\documentclass{article}

\usepackage{arxiv}

\usepackage[utf8]{inputenc} 
\usepackage[T1]{fontenc}    
\usepackage{hyperref}       
\usepackage{url}            
\usepackage{booktabs}       
\usepackage{amsfonts}       
\usepackage{nicefrac}       
\usepackage{microtype}      
\usepackage{lipsum}		    
\usepackage{graphicx}
\usepackage{authblk}
\usepackage{amsmath}
\usepackage{svg}
\usepackage{subcaption}
\usepackage{comment}
\usepackage{caption}
\usepackage{cite}
\usepackage{xcolor}

\newcommand{\bs}[1]{\boldsymbol{#1}}
\svgsetup{inkscapelatex=false}

\title{Thermodynamics-informed  super-resolution of scarce temporal dynamics data}

\author[1]{Carlos Bermejo-Barbanoj}
\author[2]{Beatriz Moya}
\author[3]{Alberto Bad\'ias}
\author[2,4]{Francisco Chinesta}
\author[1]{El\'ias Cueto}
\affil[1]{{\small ESI Group-UZ Chair of the National Strategy on Artificial Intelligence. \protect\\ Aragon Institute of Engineering Research (I3A). Universidad de Zaragoza. Zaragoza, Spain.}}
\affil[2]{{\small CNRS@CREATE LTD. Singapore.}}
\affil[3]{{\small ETSIAE, Universidad Polit\'ecnica de Madrid. Madrid, Spain.}}
\affil[4]{{\small ESI Group chair. PIMM Lab. ENSAM Institute of Technology. Paris, France. }}

\begin{document}
\maketitle

\begin{abstract}

	We present a method to increase the resolution of measurements of a physical system and subsequently predict its time evolution using thermodynamics-aware neural networks. Our method uses adversarial autoencoders, which reduce the dimensionality of the full order model to a set of latent variables that are enforced to match a prior, for example a normal distribution. Adversarial autoencoders are seen as generative models, and they can be trained to generate high-resolution samples from low-resoution inputs, meaning they can address the so-called super-resolution problem. Then, a second neural network is trained to learn the physical structure of the latent variables and predict their temporal evolution. This neural network is known as an structure-preserving neural network. It learns the metriplectic-structure of the system and applies a physical bias to ensure that the first and second principles of thermodynamics are fulfilled. The integrated trajectories are decoded to their original dimensionality, as well as to the higher dimensionality space produced by the adversarial autoencoder and they are compared to the ground truth solution. The method is tested with two examples of flow over a cylinder, where the fluid properties are varied between both examples.

\end{abstract}


\section{Introduction}
\label{sec:intro}

Resolution Augmentation techniques, frequently known as super-resolution, refer to a series of techniques that aim to enhance the level of detail of data, often an image, through computational techniques. Their main goal is to produce a high-resolution version of a low-resolution input, improving overall detail and, in some cases, revealing smaller details that may not appear in the original input. Although these techniques have been extensively studied for years by the computer vision community, the growing advances in machine learning have supposed an important boost to this field, as they allow to generate better quality high-resolution samples while improving efficiency. Deep learning based approaches, such as Convolutional Neural Networks (CNNs) \cite{LECUN_1998} and Generative Adversarial Networks (GANs) \cite{GAN, SRGAN, ESRGAN} have proved as an efficient way to augment the resolution of an image, as they are able to learn a mapping from low-resolution to high-resolution images, surpassing more traditional techniques like interpolation-based methods. One field that could benefit from these recent advances in super-resolution are predictive digital twins for physical systems. When developing a digital twin of a system, sensors are commonly employed to capture information of the fields of interest. Although sensors provide accurate measurements, usually their placement is limited to concrete areas due to physical limitations, resulting in sparse spatial measurements. Super-resolution techniques can be applied to this partial information to generate dense output fields \cite{BODE_2021, FUKAMI_2023, VINUESA_2023}.

While the Big Data paradigm has become famous in news media of all kinds, the reality is that big data is rarely available in engineering applications. Sensors are often expensive, and the storage, curation and subsequent handling of large amounts of data is no easy task. The result is that we are often faced with situations where less information is available than we would like or need. Given this situation, the application of super-resolution techniques to the world of time series forecasting becomes an urgent necessity.

The most widespread super-resolution techniques (mainly in the world of computational imaging) use black-box techniques to generate the missing information. Logically, this provides much better results than simple interpolation. However, such techniques show severe limitations, and recently new approaches to the problem have been tried, if we stick to the case of prediction of physical phenomena. Since we deal with physical phenomena, a logical and a priori very attractive option is based on taking advantage of the scientific knowledge developed over centuries of research to complement the missing information. Thus, for example, if we are faced with a fluid mechanics problem, the imposition of the Navier-Stokes equations provides valuable information to achieve a successful super-solution \cite{KELSHAW_2022}.

Predictive digital twins constitute a natural field of application of these techniques \cite{chinesta2020virtual,rasheed2019digital}. They aim to predict the time evolution of the real physical system they represent. However, these systems commonly exhibit complex behaviours which makes their real-time prediction difficult. In order to obtain a complete analysis of the phenomena that describe the behaviour of those systems, physical simulations must be performed using computational tools such Finite Element Method (FEM) for solid mechanics and Computational Fluid Mechanics (CFD) for fluid mechanics. These tools require the discretization of the domain into fine meshes, in most cases with millions of degrees of freedom. As a result, generally simulations are very computationally expensive, often making it impossible to obtain almost real-time predictions of the time evolution of the system. One approach to overcome this problem is to use model order reduction (MOR) methods, as often the solution of the system is contained in a lower-dimensional space, as stated in the manifold hypothesis \cite{FEFFERMAN_2016}. Basic approaches like Proper Orthogonal Decomposition (POD) \cite{NIROOMANDI_2008} rely on linear transformations to project the information to a low dimensional space, but they usually fail to model complex nonlinear phenomena. The ROM community has developed some techniques to overcome this limitation and obtain nonlinear mappings, like Local Linear Embedding (LLE) \cite{BADIAS_2019} and kernel Principal Component Analysis (k-PCA) \cite{BEA_KPCA}, but in recent years, deep learning based methods \cite{CODINA_2023} have been gaining popularity, with Autoencoders \cite{DEEPLEARNING} being the most common approach. Some works within the ROM community have addressed the multi-fidelity problem, for instance the Non-Intrusive Reduced Basis (NIRB) \cite{Chakir_2021, Grosjean_2023}. Those methods could benefit from the advantages of super-resolution techniques to obtain high-fidelity data or to enhance the outputs. Multi-scale problems could also benefit from the interaction of both methods, as the NIRB could handle the large scale information, while the super-resolution could refine the fine details, leading to high resolution results. In the present work we focus on autoencoders, as they have proven their capabilities to produce highly nonlinear manifolds for a wide range of applications that include physical simulations \cite{VINUESAAE,VINUESAAETRANS}. Moreover, some autoencoder architectures exhibit generative capabilities, which makes them a feasible option to generate high-resolution outputs from low-resolution data.

In order to predict the time evolution of the analysed systems, deep learning approaches can be also applied. While classically these approaches have been seen as black boxes, as they require large amounts of data and fail to generalize, leading to unreliable predictions, in the recent years there has been a growing interest in physics-consistent deep learning. These techniques consist in adding some physical knowledge of the system to neural network to guarantee the physical consistency of the solution, minimizing the amount of data needed and improving generalization capabilities. Some works in this field are based on solving the PDEs that govern the problem, which leads to very accurate results \cite{PINNS, BANERJEE_2023}. The main drawback of these methods is that they require some knowledge of the governing equations of the phenomena, and in practical applications they are often not fully known. An alternative approach is thus to enforce more general physics, or physics of a higher epistemic level. In this last case, thermodynamics comes into play as a natural choice when more detailed information is missing. Some approaches have been done by imposing the so-called GENERIC (General Equation for Non-Equilibrium Reversible-Irreversible Coupling) metriplectic structure \cite{GENERIC_I, GENERIC_II} of the problem, by means of the so-called Structure Preserving Neural Networks \cite{SPNN_QUERCUS} and Thermodynamics-Informed Graph Neural Networks \cite{TIGNN_QUERCUS}. These neural networks lead to a thermodynamically-consistent prediction that can be applied to both conservative and dissipative systems. Recently, new insights in the way we can impose the fulfillment of the first and second laws of thermodynamics to the learning process have been included in \cite{LEE_2021, GFINNS}. Previous works \cite{MOR_QUERCUS} have proved the efficiency of combining model order reduction by autoencoders and time evolution prediction by structure-preserving neural networks, leading to fast and accurate predictions.

The aim of this work is to develop a method to augment the resolution of the low-resolution fields of the state variables of a system and consequently to predict a physically-consistent evolution of this system at the high-resolution regime. The proposed methodology is very general as the used formulation to predict the time evolution of the system is valid for a wide variety of dynamical systems, although we focus on fluid mechanics. The resulting high-resolution reconstruction of the system dynamics is guaranteed to fulfill the first and second principles of thermodynamics (energy conservation and non-negative entropy production).

In this way, both the super-resolution of the state variables of our system and the prediction of the time evolution of their dynamics will be carried out under the perspective of the same formalism, the so-called GENERIC equation, whose usefulness and physical correctness for a multitude of phenomena has already been demonstrated in previous works \cite{pavelka2018multiscale}.

The structure of the paper is as follows. A description of the problem setup is presented in Section \ref{sec:prob_stat}. The methodology is presented in Section \ref{sec:methods}, where both the model order reduction autoencoder and the GENERIC formalism to predict the evolution are described. In Section \ref{sec:results} two examples are analysed: the flow past a cylinder in a Newtonian and non-Newtonian setting. Finally, the conclusions of the paper are discussed at Section \ref{sec:conclusions}.

\section{Problem statement}
\label{sec:prob_stat}

In this work we propose a framework to estimate the temporal evolution of a physical system from data, and to augment its spatial resolution, given the assumption of scarce data. We apply superresolution techniques based in the employ of deep learning and making use of the so-called dynamical system equivalence of scientific machine learning \cite{E2017}. We assume a dynamical system governed by a set of state variables \( \mbox{\boldmath$x$} \) $\in \mathcal{M} \subseteq \mathbb{R}^{D} $, with $\mathcal{M}$ the state space of these variables, assumed to evolve on a differentiable manifold in $\mathbb{R}^{D}$, thanks to the widespread manifold hypothesis \cite{FEFFERMAN_2016}. The full-order model of a physical phenomenon can be expressed as a system of differential equations that give the temporal evolution of a set of state variables \( \mbox{\boldmath$x$} \),
\begin{equation} \label{eq:general_eq}
	\mbox{\boldmath$\dot{x}$} = \frac{d\mbox{\boldmath$x$}}{dt} = \bs F \left( \mbox{\boldmath$x$}, t \right), \: t \text{ in }  \mathcal{I} = \left( 0, T \right], \: \mbox{\boldmath$x$} \left( 0 \right) =  \mbox{\boldmath$x$}_{0},
\end{equation}
where $t$ refers to the time coordinate in the time interval $\mathcal{I}$ and  $\bs F \left( \mbox{\boldmath$x$}, t \right)$ is an a priori unknown nonlinear function that represents the flow map of the governing variables. The identification of this function $\bs F$ from data is precisely the objective of this work, where we assume that we work in a scarce data scenario.

Since in the most common applications of such techniques (such as the aforementioned digital twins) there is the additional circumstance of strong real-time constraints, it is also assumed that there is a need to work on reduced models of the physics under study. The dimensionality reduction procedure looks for a simpler representation of the full-order state vector represented by \( \mbox{\boldmath$x$} \), through a set of reduced (also denoted as latent in the literature of machine learning) variables \( \mbox{\boldmath$z$} \) $\in \mathcal{N} \subseteq \mathbb{R}^{d} $ contained in a manifold with a dimensionality lower than the original space $\mathcal{M}$. The mapping between both spaces is denoted by $\phi : \mathcal{M} \subseteq \mathbb{R}^{D} \rightarrow \mathbb{R}^{d}$,  where $d \ll D$. An inverse mapping $\phi^{-1}$ allows to undo the transformation, recovering the information in the full-order space

The goal of this paper is two-fold. First, to find a mapping $\phi$ for a dynamical system governed by Eq. (\ref{eq:general_eq}) that allows us to predict its temporal evolution under stringent real-time constraints on a reduced-order manifold, and then to augment the spatial dimensionality of the data, back to the original, full-order state space manifold. The mapping $\phi$ allows to learn the underlying physics of the system in the reduced space $\mathcal{N}$ and then predict its temporal evolution. In order to obtain a physically consistent prediction of the system, the solution must fulfill the laws of thermodynamics, which are enforced by assuming that their evolution occurs under the GENERIC framwework. The second objective is to achieve this while simultaneously augmenting the spatial resolution of the data to generate a dense solution field, thus obtaining a solution space with a higher dimensionality than the  original space $\mathcal{M}$.

\section{Methodology}
\label{sec:methods}

The proposed framework splits the problem in two main steps. First, the low-resolution, full order model is encoded (projected) onto a reduced-order manifold (or latent space) with an autoencoder, thus achieving a nonlinear mapping $\phi$. The autoencoder learns a coded representation of the physical system, which allows to work with the data in a compact form. Moreover, the autoencoder is trained to generate high resolution fields of the state variables from the low resolution input data.

\begin{figure}[h]
    \centering
    \includegraphics[width=\textwidth]{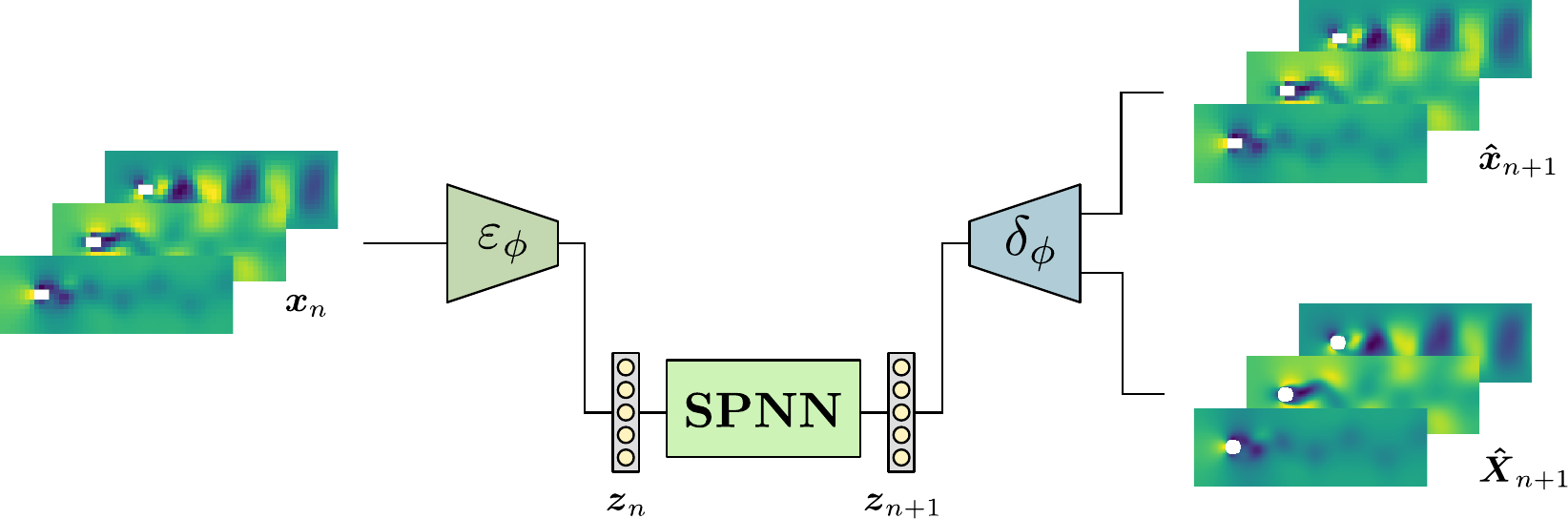}
    \caption{Scheme of the proposed framework. First, an encoder is used to reduce the dimensionality of the problem, obtaining a set of reduced variables or latent code. Then, a structure-preserving neural network (SPNN) is trained to integrate the time evolution of the reduced variables of the system. Thus, given the state of the system at time instant $n$, the net obtains the state at time instant $n + \Delta n$. Finally the decoder is used to recover the data to its original dimensionality and to generate the output in a resolution that is higher that the input one.}
    \label{fig:global_scheme}
\end{figure}

Then, a structure-preserving neural network is trained with (low resolution, but full order) simulation data so as to obtain a temporal prediction of the evolution of the dynamics of the system. This network predicts the time evolution of the system by using the GENERIC formalism. Finally, these latent variables are projected back by the decoder to both the original manifold of the low-resolution full order model and a higher resolution manifold. A general scheme of this procedure is shown in Fig. \ref{fig:global_scheme}.
In this work, the full order model data has been generated in silico, although this procedure could be applied to measurements coming from a real physical system.

\subsection{Model reduction with Adversarial Autoencoders}
\label{subsec:aae}

An autoencoder is a type of neural network that learns an efficient codification or embedding of data. This results in a dimensionality reduction of the input information into a set of latent variables, which ideally contain the same information as the original data. The classical autoencoder architecture is composed by two basic elements: an encoder, $\varepsilon_{\phi}$, that maps the high-dimensional information  into a low-dimensional code,  and a decoder, $\delta_{\phi}$, that applies the inverse operation, recovering the information into the original full-order manifold.
\begin{equation} \label{eq:ae_encoder}
	 \varepsilon_{\phi}: \mathbb{R}^{D} \rightarrow \mathbb{R}^{d}, \; \; \mbox{\boldmath$z$} = \varepsilon_{\phi}\left( \mbox{\boldmath$x$} \right),
\end{equation}
\begin{equation} \label{eq:ae_decoder}
	\delta_{\phi}: \mathbb{R}^{d} \rightarrow \mathbb{R}^{D}, \; \; \mbox{\boldmath$\hat{x}$} = \delta_{\phi}\left( \mbox{\boldmath$z$} \right).
\end{equation}
In this work, we use adversarial autoencoders (AAEs) \cite{AAE}. This kind of autoencoder enforces the latent vector to follow a desired distribution or prior, similarly to variational autoencoders (VAEs) \cite{VAE}, but instead of predicting the mean and standard deviation to enforce that the latent code follows the prior, it is enforced by using an additional network called discriminator. This allows the latent variables to follow not only normal distributions, like VAEs, but also more complex ones.

The discriminator is a simple neural network, usually a Multilayer Perceptron (MLP), that takes as input the latent code generated by the autoencoder and a random sample that follows the prior. It compares both to determine how close the latent code is to the prior. As the training process advances, the latent code produced by the AAE is closer to the prior, which means that the discriminator finds harder to discern if the sample comes from the prior or from the autoencoder.

\begin{figure}[h]
    \centering
    \includegraphics[width=0.9\textwidth]{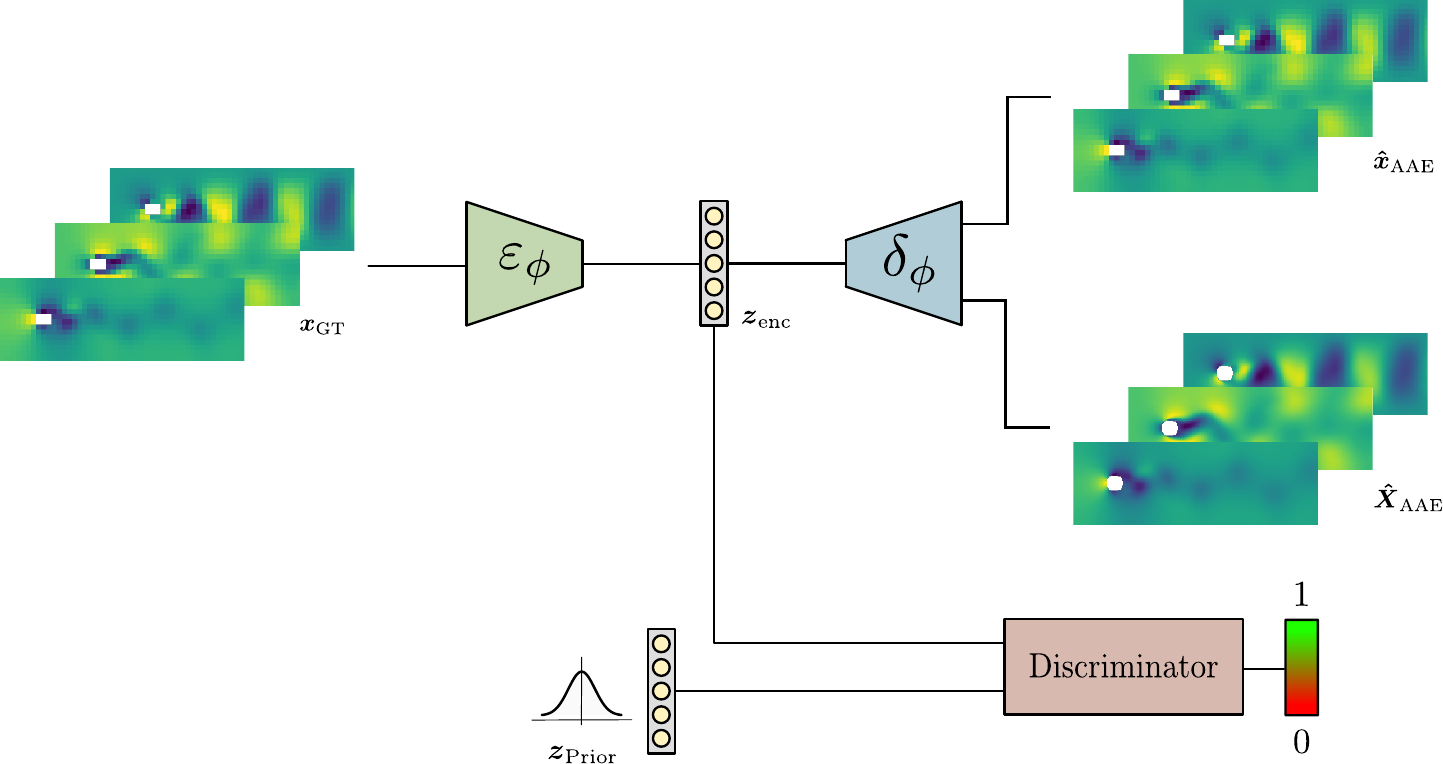}
    \caption{Scheme of the adversarial autoencoder (AAE). The autoencoder takes a snapshot of the simulation as input and learns an encoded representation of the data. The decoder recovers the data to its original dimensionality, \( \mbox{\boldmath$\hat{x}$}_{\text{AAE}}\), and it is trained to generate a higher-resolution output, \( \mbox{\boldmath$\hat{X}$}_{\text{AAE}}\), achieving "superresolution" of the analysed simulation. The discriminator takes as input a sample that follows the proposed distribution or prior (in this work a normal distribution) and the latent code generated by the autoencoder. The discriminator output is in the range between 0 and 1. An output close to 0 means that the latent code does not match the prior distribution, while if the output is close to 1 the autoencoder is able to produce a latent code that fits into the prior distribution.}
    \label{fig:aae_scheme}
\end{figure}

AAEs are seen as a mix between VAEs and Generative Adversarial Neural Networks (GANs) \cite{GAN}, as they enforce the latent code to follow a prior  but  make use of a discriminator to ensure that this prior is matched. Like VAEs and GANs, AAEs are considered as generative models. This results in a very useful feature for the proposed task, as they can be used to generate a high-resolution output from a low-resolution input. The resolution augmentation has been achieved by training the decoder of the AAE to output the low-dimensional data (same as input, as in classical autoencoders) and also the high-dimensional data, supervising the training with the ground truth information. The AAE scheme can be seen in Fig. \ref{fig:aae_scheme}.

The loss function of the Autoencoder is composed therefore by three terms:

\begin{itemize}
	\item \textbf{Low-Resolution data loss}: The output of the autoencoder, 
	$\hat{\bs x}$, must match the ground truth, in this case the input of the network, the low resolution pressure and velocity fields, 
	$\bs x$. The accuracy of the network is evaluated using the mean squared error:
\begin{equation} \label{eq:aae_loss_data_low} 		\mathcal{L}_{\text{mse, LR}}^{\text{AAE}} = \frac{1}{\tt n_{\text{snap}}}\sum_{i=0}^{\tt n_{\text{snap}}} \left( 
\bs x_i-\hat{\bs x}_i\right)^2,
	\end{equation}
where the subscript $i$ refers to the snapshot number, $i=1, \ldots, \tt n_{\text{snap}}$.	
	\item \textbf{High-Resolution data loss}: The output of the autoencoder, 
	$\hat{\bs X}$, must match the ground truth, the high-resolution pressure and velocity fields obtained from the in-silico simulations, 
	$\bs X$. As with the low resolution data, the accuracy of the autoencoder is evaluated using the mean squared error:	
		\begin{equation} \label{eq:aae_loss_data_high}
		\mathcal{L}_{\text{mse, HR}}^{\text{AAE}} = \frac{1}{\tt n_{\text{snap}}}\sum_{i=0}^{\tt n_{\text{snap}}} \left( 
		\bs X_i-\hat{\bs X}_i\right)^2.
	\end{equation}
	
	\item \textbf{Adversarial loss}: The third term of the loss function is the contribution of the discriminator, $\mathcal{L}_{\text{adv}}^{\text{AAE}}$. This term measures the likehood between the proposed distribution and the distribution obtained by the encoder.
\end{itemize}

The final loss function of the autoencoder is composed of a weighted sum of all the terms. A hyperparameter, $\lambda_{\text{adv}}^{\text{AAE}}$ is added to control its influence to the total loss function,
\begin{equation} \label{eq:aae_loss}
	\mathcal{L}^{\text{AAE}} = \mathcal{L}_{\text{rec}}^{\text{AAE}} + \lambda^{\text{AAE}}_{\text{adv}}\cdot \mathcal{L}_{\text{adv}}^{\text{AAE}},
\end{equation}
where the $\mathcal{L}_{\text{rec}}^{\text{AAE}}$ term represents the reconstruction capabilities of the autoencoder and is composed by the terms associated to the low-resolution and high-resolution fields,
\begin{equation} \label{eq:aae_loss_reconst}
	\mathcal{L}_{\text{rec}}^{\text{AAE}} = \mathcal{L}_{\text{mse, LR}}^{\text{AAE}} + \mathcal{L}_{\text{mse, HR}}^{\text{AAE}}.
\end{equation}

\subsection{Learning the dynamical evolution of the system by Structure-Preserving Neural Network}
\label{subsec:predicting_dynamics}

One of our main interests is to develop a framework that satisfies a priori, by construction, known principles of physics about the phenomenon at hand. This is crucial, as we want our framework to provide credible, robust and accurate predictions to help in fields like decision-making, and this can only be achieved with predictions that fulfill the basic principles of physics. In our approach, this is achieved by using physics principles as inductive bias. An inductive bias is a set of assumptions about the data that prioritise one solution over the rest---precisely, the one fulfilling known physical principles---, preventing the learning process from finding a local minimum of the loss function.

Maybe the most popular method in our community at this moment is the so-called Physics-Informed Neural Networks (PINN) \cite{PINNS}, in which we enforce the fulfillment of a particular partial differential equation that governs our system. However, there are some situations where the governing equations are not well known or they cannot be applied easily. In other situations, there are models that are well known but nevertheless provide unconvincing results in predicting the evolution of the system. In this case, a very attractive option is to learn only the "ignorance" about the physical behaviour, so that the prediction is the sum of the evolution predicted by the model and the prediction of the learnt ignorance model about the system. This is the approach that has been followed, for example, in \cite{chinesta2020virtual,moya2022digital}.

For that reason, we want to guarantee the physical meaning of the solution, but without enforcing any particular physical equation.

For this purpose, we use a structure-preserving neural network (SPNN) \cite{SPNN_QUERCUS}. Structure-preserving neural networks refer to a class of methods that are constructed to satisfy some high-level epistemic properties of the problem, for example, the principles of thermodynamics. SPNN can be applied to conservative and dissipative problems, ensuring that the principles of thermodynamics are satisfied by construction. This property allows us to use the thermodynamics laws as an inductive bias \cite{BATTAGLIA_2018}, ensuring the physical consistency of the results. 

\subsubsection{GENERIC Formalism}
\label{subsubsec:GENERIC}

To guarantee the physical meaning of the solution, we enforce the "General Equation for Non-Equilibrium Reversible-Irreversible Coupling", usually referred as GENERIC formalism \cite{GENERIC_I, GENERIC_II}. This formalism is a generalization of the classic Hamiltonian formulation to dissipative systems. This approach assumes the reversible or conservative contribution to be of Hamiltonian form, thus requiring an energy function and a Poisson bracket. The irreversible contribution to the energetic balance is generated by the non-equilibrium entropy and an irreversible or friction bracket \cite{MORRISON_METRIPLECTIC}.

The GENERIC formulation of time evolution for non-equilibrium systems, parameterised by a set of state variables able to describe the evolution of the energy of the system, \( \mbox{\boldmath$z$} \)---the choice is thus not unique---, is given by:
\begin{equation} \label{eq:bracket_form}
	\frac{d\mbox{\boldmath$z$}}{dt} = \{ \mbox{\boldmath$z$}, E \} + \left[ \mbox{\boldmath$z$}, S \right],
\end{equation}
where the so-called Poisson bracket $\{ \cdot,\cdot\}$ and dissipative bracket $[\cdot,\cdot]$ have been used.
For practical use, the bracket notation is often reformulated using two linear operators:
\begin{equation} \label{eq:reformulation}
	\mbox{\boldmath$L$}:T^{*}\mathcal{M} \rightarrow T\mathcal{M}, \; \mbox{\boldmath$M$}:T^{*}\mathcal{M} \rightarrow T\mathcal{M},
\end{equation}
where $T^{*}\mathcal{M}$ and $T\mathcal{M}$ represent, respectively, the cotangent and tangent bundles of the state space $\mathcal{M}$. The operator \( \mbox{\boldmath$L$} \) represents the Poisson bracket and must be skew-symmetric, while the operator \( \mbox{\boldmath$M$} \), the friction matrix, describes the irreversible part of the system and must be positive semidefinite to make sure that the dissipation rate is positive. For phenomena involving plasticity, for instance, this approach may not be valid, and a more general form of the dissipative term should be considered. A more general one is developed in \cite{Mielke,mielke2011formulation}, among other references.

Replacing the original bracket formulation in Eq. (\ref{eq:bracket_form}) with their respective operators, the time evolution equation for the state variables \( \mbox{\boldmath$z$} \) is derived,
\begin{equation} \label{eq:generic}
	\frac{d\mbox{\boldmath$z$}}{dt} = \mbox{\boldmath$L$}\frac{\partial E}{\partial \mbox{\boldmath$z$}} + \mbox{\boldmath$M$}\frac{\partial S}{\partial \mbox{\boldmath$z$}}.
\end{equation}

The equation is completed by adding the so-called degeneracy conditions:
\begin{equation} \label{eq:degeneration_bracket}
	\{ S, \mbox{\boldmath$z$} \} = \left[ E, \mbox{\boldmath$z$} \right] = \mbox{\boldmath$0$}.
\end{equation}
The first expression states that the entropy is a degenerate functional of the Poisson bracket, and shows the reversible nature of the Hamiltonian contribution to the dynamics. The second expression states that the energy is a degenerate functional of the friction matrix, so the total energy of the system is conserved. These conditions can be reformulated into a matrix form in terms of the previously defined \( \mbox{\boldmath$L$} \) and \( \mbox{\boldmath$S$} \) operators, which results in the following degeneracy conditions: 
\begin{equation} \label{eq:degeneration_generic}
	\mbox{\boldmath$L$}\frac{\partial S}{\partial \mbox{\boldmath$z$}} = \mbox{\boldmath$M$}\frac{\partial E}{\partial \mbox{\boldmath$z$}} = \mbox{\boldmath$0$}.
\end{equation}
The degeneracy conditions, in addition to the non-negativeness of the irreversible bracket, guarantees that the first (energy conservation) and the second (entropy inequality) laws of thermodynamics are fulfilled.
\begin{equation} \label{eq:thermodynamics_fullfilled}
	\frac{dE}{dt} = 0, \; \frac{dS}{dt} \geq 0.
\end{equation}

\subsubsection{Structure-Preserving Neural Networks}
\label{subsubsec:spnn}
The structure-preserving neural networks impose the GENERIC formalism to guarantee the thermodynamical consistency of the solution. In order to work with the data coming from the simulation, the GENERIC formalism is discretized along time intervals $\Delta t$,
\begin{equation} \label{eq:generic_discrete}
	\frac{{\mbox{\boldmath$z$}}_{n+1} - {\mbox{\boldmath$z$}}_{n}}{\Delta t} = \mathsf{L}_{n}\Big( \frac{\mathsf{D} E}{\mathsf{D} \bs z} \Big)_n + \mathsf{M}_{n}\Big( \frac{\mathsf{D} S}{\mathsf{D} \bs z} \Big)_n,
\end{equation}
where we employ the subscript $n$ to refer to time instant $t=n\Delta t$ and, therefore, $n+1$ to refer to $t+\Delta t=(n+1)\Delta t$.

In this scheme the time derivative is substituted by a forward-Euler scheme with time increments $\Delta t$. The accuracy and stability of different time discretisations of the GENERIC equation have been deeply analysed in \cite{romero2010algorithms,romero2010algorithms2}.The Poisson and friction operators are discretized as $\mathsf{L}_{n}$ and $\mathsf{M}_{n}$. Similarly, energy and entropy gradients are discretized as $\Big( \frac{\mathsf{D} E}{\mathsf{D} \bs z} \Big)_n$ and $\Big( \frac{\mathsf{D} S}{\mathsf{D} \bs z} \Big)_n$. Eq.(\ref{eq:generic_discrete}) can be rewritten to the proposed integration scheme to predict the temporal evolution of the system:
\begin{equation} \label{eq:generic_integration}
	\mbox{\boldmath$z$}_{n+1} = \mbox{\boldmath$z$}_{n} + {\Delta t} \cdot \left( \mathsf{L}_{n}\Big( \frac{\mathsf{D} E}{\mathsf{D} \bs z} \Big)_n + \mathsf{M}_{n}\Big( \frac{\mathsf{D} S}{\mathsf{D} \bs z} \Big)_n \right) .
\end{equation}
Additionally, discretized degeneracy conditions are added to ensure the thermodynamical consistency of the prediction:
\begin{equation} \label{eq:degeneration_discrete}
	\mathsf{L}_{n}\Big( \frac{\mathsf{D} S}{\mathsf{D} \bs z} \Big)_n = \bs 0, \; \mathsf{M}_{n}\Big( \frac{\mathsf{D} E}{\mathsf{D} \bs z} \Big)_n = \bs 0.
\end{equation}

The GENERIC structure is imposed to the encoded space learnt by the adversarial autoencoder, similarly to \cite{MOR_QUERCUS, BEA_MOR}. The SPNN is a feed-forward neural network composed by a set of fully connected layers. The input of the net is the encoded state vector at a given given timestep $ \mbox{\boldmath$z$}_{n}^{\text{AAE}}$. The output from the net is a vector containing the predicted $\mathsf{L}_{n}$ and $\mathsf{M}_{n}$ matrices, as well as the predicted energy and entropy gradients, $\Big( \frac{\mathsf{D} E}{\mathsf{D} \bs z} \Big)_n$ and $\Big( \frac{\mathsf{D} S}{\mathsf{D} \bs z} \Big)_n$. 

Actually, to enforce the skew-symmetry and positive semi-definiteness of matrices $\bs{L}$ and $\bs{M}$, the output of the network is a pair of matrices $\bs{l}$ and $\bs{m}$, reshaped in lower-triangular matrices,
\begin{equation}
\label{eq:gnn_output}
\bs{L}=\bs{l}-\bs{l}^\top, \qquad \bs{M}=\bs{m}\bs{m}^\top. 
\end{equation}

Then, using the integration scheme showed in Eq.(\ref{eq:generic_integration}), the reduced space state vector at the next time step is obtained $ \mbox{\boldmath$z$}_{n+1}^{\text{SPNN}}$.

\begin{figure}[h]
    \centering
    \includegraphics[width=0.8\textwidth]{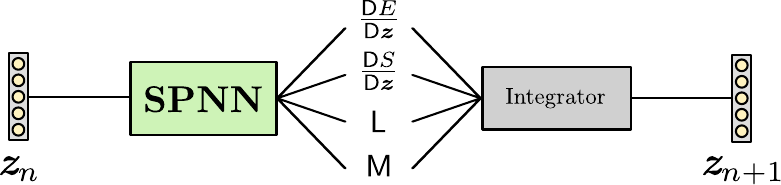}
    \caption{Scheme of the structure-preserving neural network (SPNN). The SPNN is trained to predict the full time evolution of the latent variables generated by the AAE by applying the GENERIC structure of the underlying physics of the problem. The network takes the current snapshot as input and outputs the L and M matrices, as well as the energy and entropy gradients. Then, they are integrated following the GENERIC formalism, as shown in Eq. \ref{eq:generic_integration}, and the latent variables of the next snapshot are obtained. This process can be done iteratively, obtaining the rollout prediction of the full simulation.}
    \label{fig:spnn_scheme}
\end{figure}

The loss function used to train the SPNN is composed by two different terms:
\begin{itemize}
	\item \textbf{Data loss}: The output of the integration scheme, $ \mbox{\boldmath$z$}_{n+1}^{\text{SPNN}}$, must match the ground truth, in this case the encoded state vector, $ \mbox{\boldmath$z$}_{n+1}^{\text{AAE}}$, predicted by the autoencoder. The accuracy of the network is evaluated using the mean squared error:
	\begin{equation} \label{eq:spnn_loss_data}
		\mathcal{L}_{\text{mse}}^{\text{SPNN}} = \frac{1}{\tt n_{\text{snap}}}\sum_{n=0}^{\tt n_{\text{snap}}} \left( \mbox{\boldmath$z$}_{n+1}^{\text{AAE}} - \mbox{\boldmath$z$}_{n+1}^{\text{SPNN}} \right).
	\end{equation}
	\item \textbf{Degeneracy conditions loss}: The loss function includes the fulfilment of the degeneracy conditions, ensuring the thermodynamical consistency of the solution. They are measured as the sum of the squared values of both conditions:
	\begin{equation} \label{eq:spnn_loss_degeneration}
		\mathcal{L}_{\text{degen}}^{\text{SPNN}} = \frac{1}{\tt n_{\text{snap}}}\sum_{n=0}^{\tt n_{\text{snap}}} \left(  \left( \mathsf{L}_{n}\Big( \frac{\mathsf{D} S}{\mathsf{D} \bs z} \Big)_n \right)^{2} + \left( \mathsf{M}_{n}\Big( \frac{\mathsf{D} E}{\mathsf{D} \bs z} \Big)_n \right)^{2} \right).
	\end{equation}
\end{itemize}
The final loss function is composed of a weighted sum of both terms. A hyperparameter $\lambda_{\text{mse}}^{\text{SPNN}}$ is added to control its influence and balance both of them,
\begin{equation} \label{eq:spnn_loss}
	\mathcal{L}^{\text{SPNN}} = \lambda_{\text{mse}}^{\text{SPNN}} \cdot \mathcal{L}_{\text{mse}}^{\text{SPNN}} + \mathcal{L}_{\text{degen}}^{\text{SPNN}}.
\end{equation}

\section{Results}
\label{sec:results}

\subsection{Example 1: Flow past a cylinder of a Newtonian fluid}
\label{subsec:example1}

\subsubsection{Database generation}
\label{subsubsec:data1}

The first example consists in an unsteady flow past a cylindrical obstacle. The geometry of the obstacle is fixed for all examples and the flow conditions are varied by modifying the freestream velocity, which results in a variable Reynolds regime and generates a Kármán vortex street that exhibits a periodic behaviour during the steady state.
The state variables for the flow past a cylinder are the velocity and pressure fields,
\begin{equation} \label{eq:state_var_non_newt}
	\mathcal{S} = \lbrace \mbox{\boldmath$x$} = \left( \mbox{\boldmath$u$},P \right) \in \mathbb{R}^{2} \times \mathbb{R} \rbrace.
\end{equation}
The ground truth simulations are computed solving the 2D Navier-Stokes equations using OpenFOAM software \cite{OPENFOAM}. No-slip condition is applied in the cylinder obstacle . The fluid is assumed to have a Newtonian behaviour with density of $\rho = 1$ and dynamic viscosity $\mu = 10^{-3}$. The freestream velocity is contained within the interval \( \mbox{\boldmath$u$} \in \left[ 0.8, 3.4 \right] \), resulting in a total of $N_{\text{sim}} = 27$ cases. Each case is discretized in ${\tt n}_{\text{snap}}= 800$ time increments of $\Delta t = 0.005$.

\begin{figure}[h]
	\centering
		\begin{subfigure}{0.49\textwidth}
			\includegraphics[width=\textwidth]{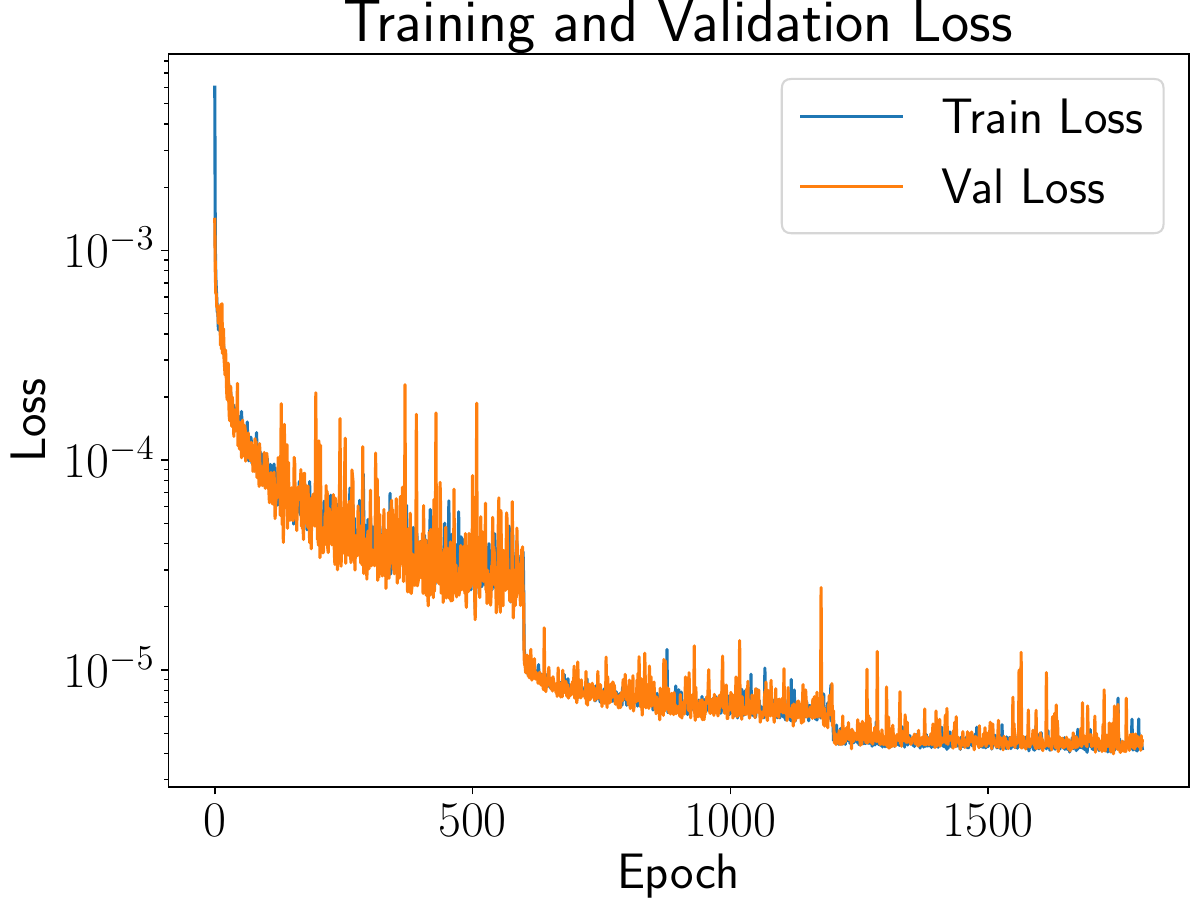}
            \caption{AAE Loss}
			\label{fig:aae_loss_newt}
	   \end{subfigure}
	   \begin{subfigure}{0.49\textwidth}
			\includegraphics[width=\textwidth]{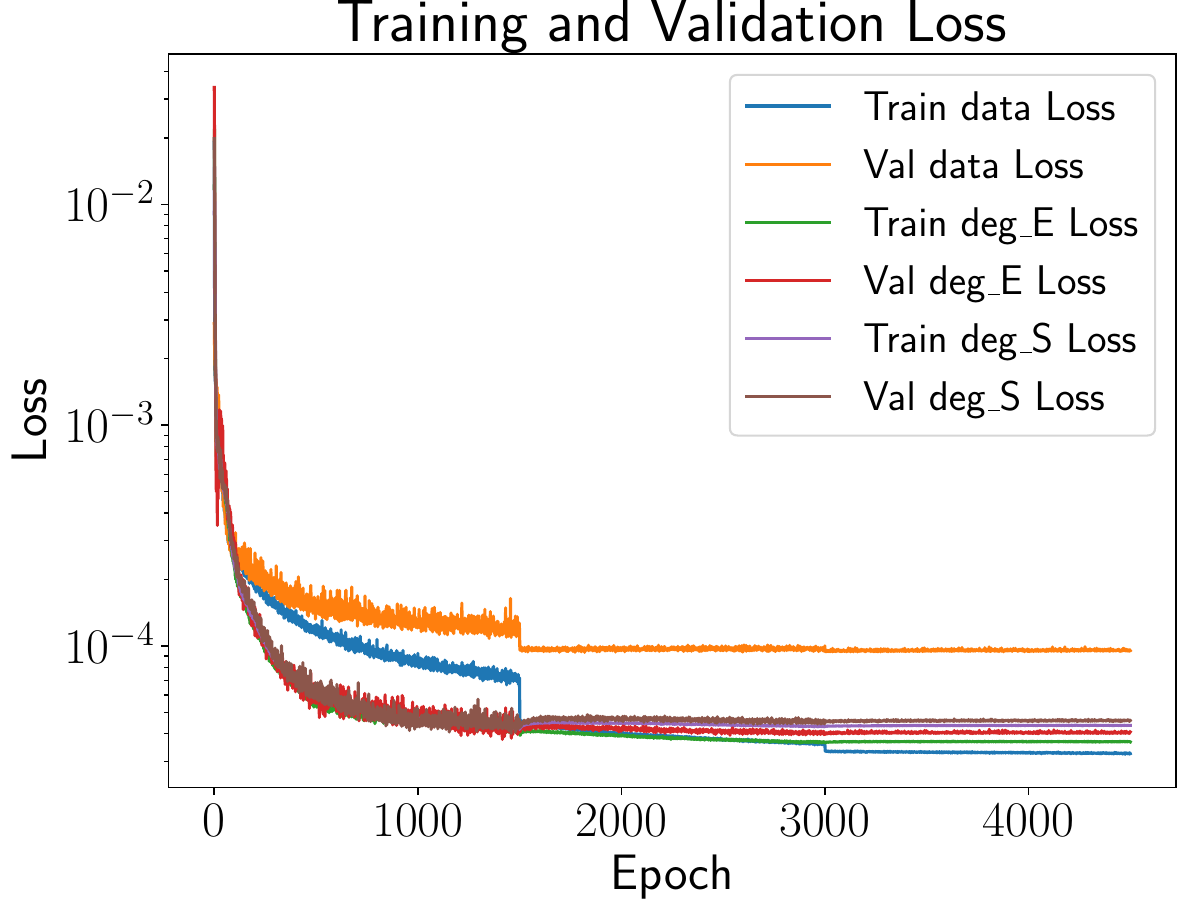}
            \caption{SPNN Loss}
			\label{fig:spnn_loss_newt}
	    \end{subfigure}
	\caption{Training and validation loss curves for the Adversarial Autoencoder (Left) and the SPNN (right).}
	\label{fig:loss_newt}
\end{figure}

The input of the autoencoder are the low resolution velocity and pressure fields, with size $3 \times 16 \times 48$ , while the output are the velocitiy (two components) and pressure (a scalar) fields at the original resolution and a higher one, with sizes  $3 \times 16 \times 48$ and $3 \times 64 \times 192$, respectively. Both the encoder and decoder use convolutional layers with $ N_{\text{ch}} = 64$ channels and a kernel size of $k = 3$, following a ResNet-like structure \cite{RESNET}. The number of latent variables at the bottleneck is set to $d = 5$. The activation function used is the Leaky-ReLU with a negative slope of $0.1$, except for the last layer of both the encoder and decoder, where linear activations are used. The adversarial hyperparameter weight is set to $\lambda_{\text{adv}}^{\text{AAE}} = 10^{-3}$. The optimizer used is Adam \cite{ADAM} with a learning rate set to $l_{r}^{\text{AAE}} = 10^{-4}$ with decreasing order of magnitude on epochs 600 and 1200, a weight decay set to $w_{d}^{\text{AAE}} = 10^{-4}$, and a total number of $N_{\text{epochs}} = 1800$ epochs.  Latent variables obtained at the bottleneck are then used as input variables for the structure preserving neural network that, as explained before, operated in the latent manifold. The training and validation loss curves for the autoencoder are shown in Fig. \ref{fig:aae_loss_newt}

The SPNN input size coincides with the AAE latent dimension, $N_{\text{in}}^{\text{SPNN}} = d = 5$, while the output size is $N_{\text{out}}^{\text{SPNN}} = d \cdot \left( d + 2 \right) = 35$, see Eq. (\ref{eq:gnn_output}). The number of hidden layers of the SPNN is $N_{\text{hl}}^{\text{SPNN}}=5$ with 100 neurons each one, Leaky-ReLU activations and linear for the last layer. The data weight hyperparameter is set to $\lambda_{\text{data}}^{\text{SPNN}} = 10^{2}$. The SPNN is trained for $N_{\text{epochs}} = 4500$ epochs using the Adam optimizer. The learning rate is set to  $l_{\text{r}}^{\text{SPNN}} = 10^{-3}$, decreasing one order of magnitude on epoch 1500 and 3000. The weight decay is set to $w_{\text{d}}^{\text{SPNN}} = 10^{-4}$ and noise variance is added to the train set, $\sigma_{\text{noise}}^{2} = 10^{-6}$. The Fig. \ref{fig:spnn_loss_newt} shows the training and validation curves for the SPNN.

\subsubsection{Results}
\label{subsubsec:results_ex1}

\begin{figure}[h]
	\centering
		\begin{subfigure}{0.49\textwidth}
			\includegraphics[width=\textwidth]{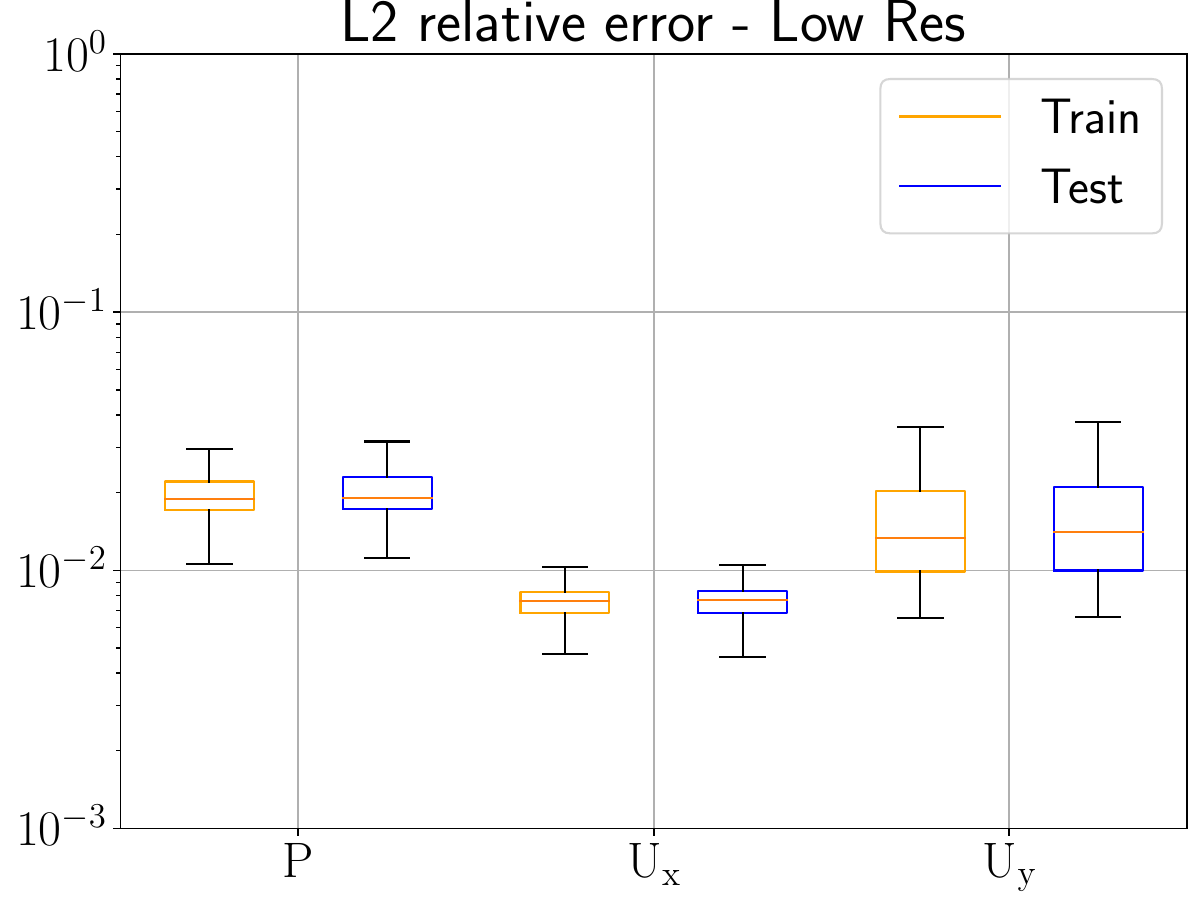}
			\label{fig:relative_error_aae_newt_lr}
	   \end{subfigure}
	   \begin{subfigure}{0.49\textwidth}
			\includegraphics[width=\textwidth]{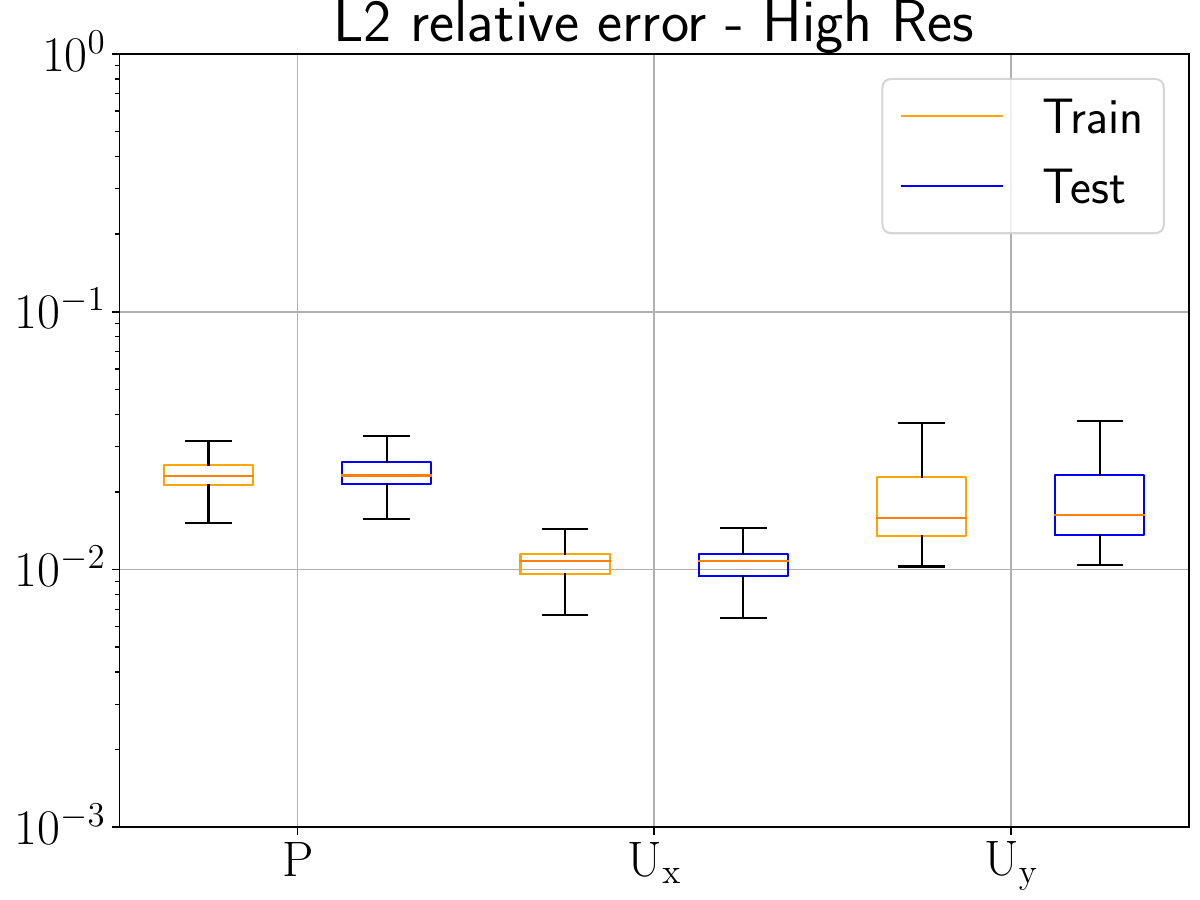}
			\label{fig:relative_error_aae_newt_hr}
	    \end{subfigure}
	\caption{Box plots for the relative L2 error of the autoencoder for all the snapshots of the newtonian fluid for both train and test cases, in low (left) and high (right) resolution. The state variables represented are pressure (P) and velocity ($U_{x}$ and $U_{y}$)}
	\label{fig:aae_newt_relative_error}
\end{figure}

Fig. \ref{fig:aae_newt_predictions} shows the prediction achieved for the pressure and velocity fields predicted by the autoencoder in low and high resolution, as well as the absolute error for each field. The AAE prediction shows good agreement with the reconstructed low resolution fields and those generated in high resolution. Fig. \ref{fig:aae_newt_relative_error} shows a box plot of the data error for the train and test sets, obtaining a mean error lower than 3\% for the pressure and velocity fields in both low and high resolution.

\begin{figure}[h]
	\centering
		\begin{subfigure}{\textwidth}
			\begin{subfigure}{\textwidth}
				\includegraphics[width=\textwidth]{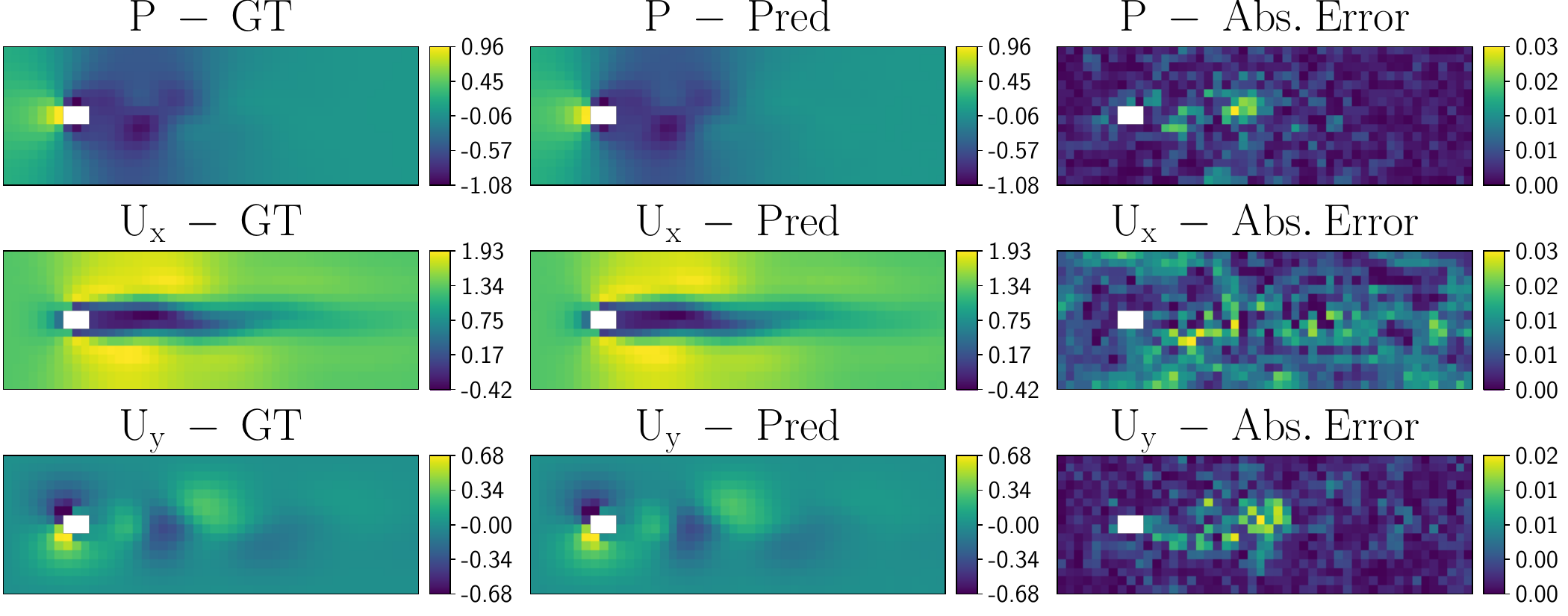}
				\caption{Low resolution}
				\label{fig:aae_example_1_newt_lr}
	   		\end{subfigure}
	   		\vfill
	   		\begin{subfigure}{\textwidth}
				\includegraphics[width=\textwidth]{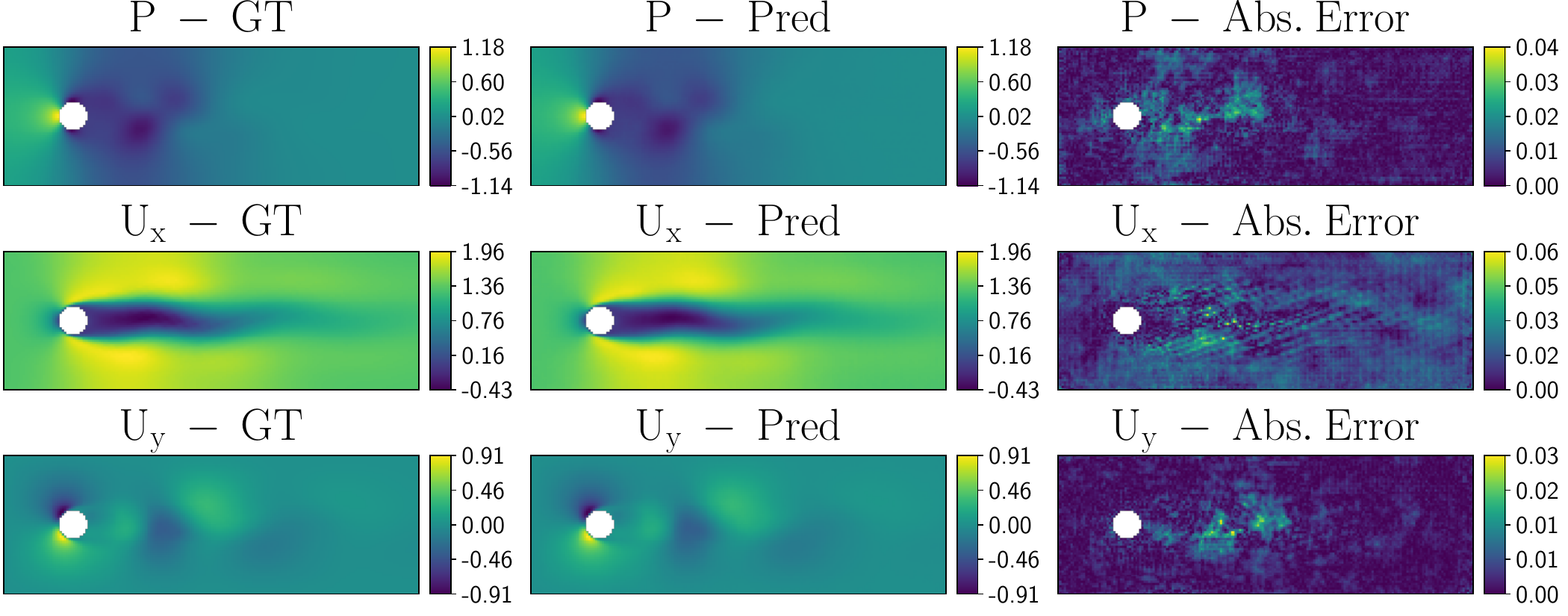}
				\caption{High resolution}
				\label{fig:aae_example_1_newt_hr}
	   		\end{subfigure}
	   \end{subfigure}
	\caption{Results of the prediction made by the Adversarial Autoencoder (AAE). \ref{fig:aae_example_1_newt_lr}: Low resolution Ground Truth (GT), AAE prediction and absolute error for $P$, $U_{x}$ and $U_{y}$. The ground truth is the input of the AAE and it predicts the low resolution fields and the high resolution fields shown in Fig. \ref{fig:aae_example_1_newt_hr}. \ref{fig:aae_example_1_newt_hr}: High resolution GT, AAE prediction and absolute error fields for the same snapshot shown in Fig. \ref{fig:aae_example_1_newt_lr}. \ref{fig:aae_example_2_newt_lr}: Low resolution fields for a different snapshots. \ref{fig:aae_example_2_newt_hr}: High resolution GT, AAE pred. and abs. error fields for the snapshot shown in Fig. \ref{fig:aae_example_2_newt_lr}.}
	\label{fig:aae_newt_predictions}
\end{figure}
\begin{figure}[h]\ContinuedFloat
	\centering
		\begin{subfigure}{\textwidth}
			\begin{subfigure}{\textwidth}
				\includegraphics[width=\textwidth]{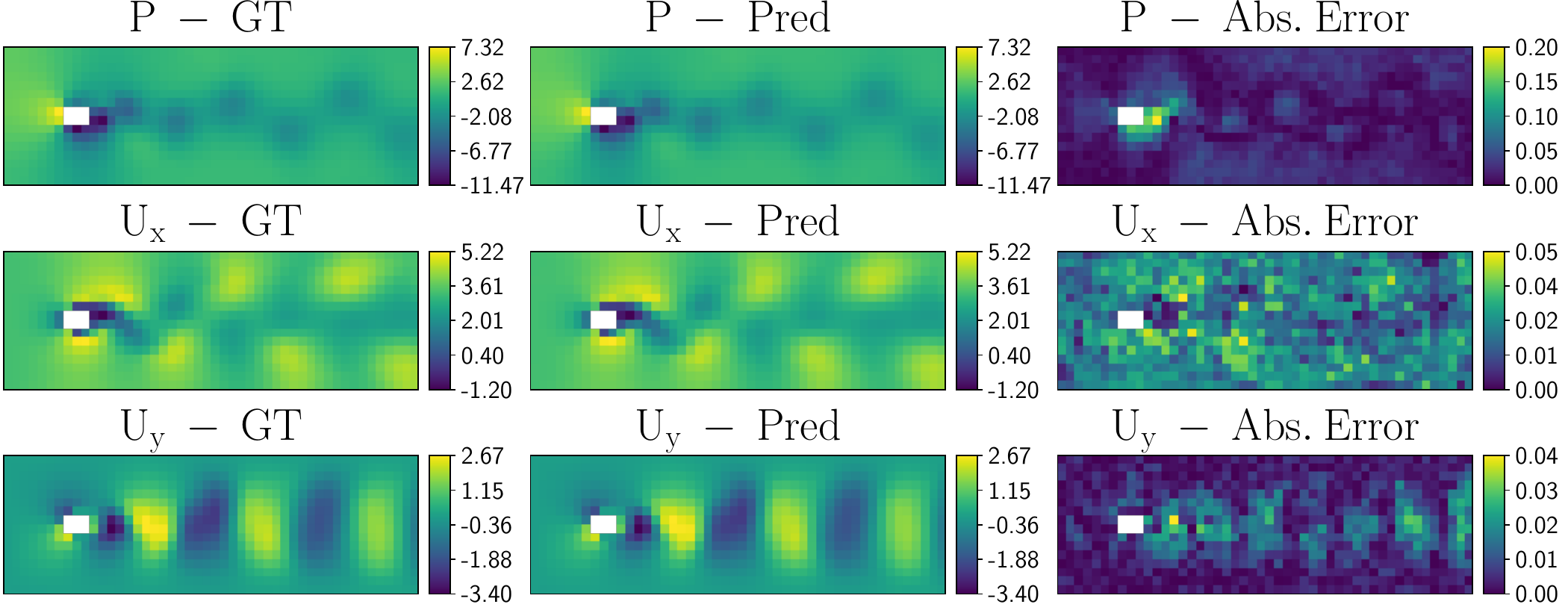}
				\caption{Low resolution}
				\label{fig:aae_example_2_newt_lr}
	   		\end{subfigure}
	   		\vfill
	   		\begin{subfigure}{\textwidth}
				\includegraphics[width=\textwidth]{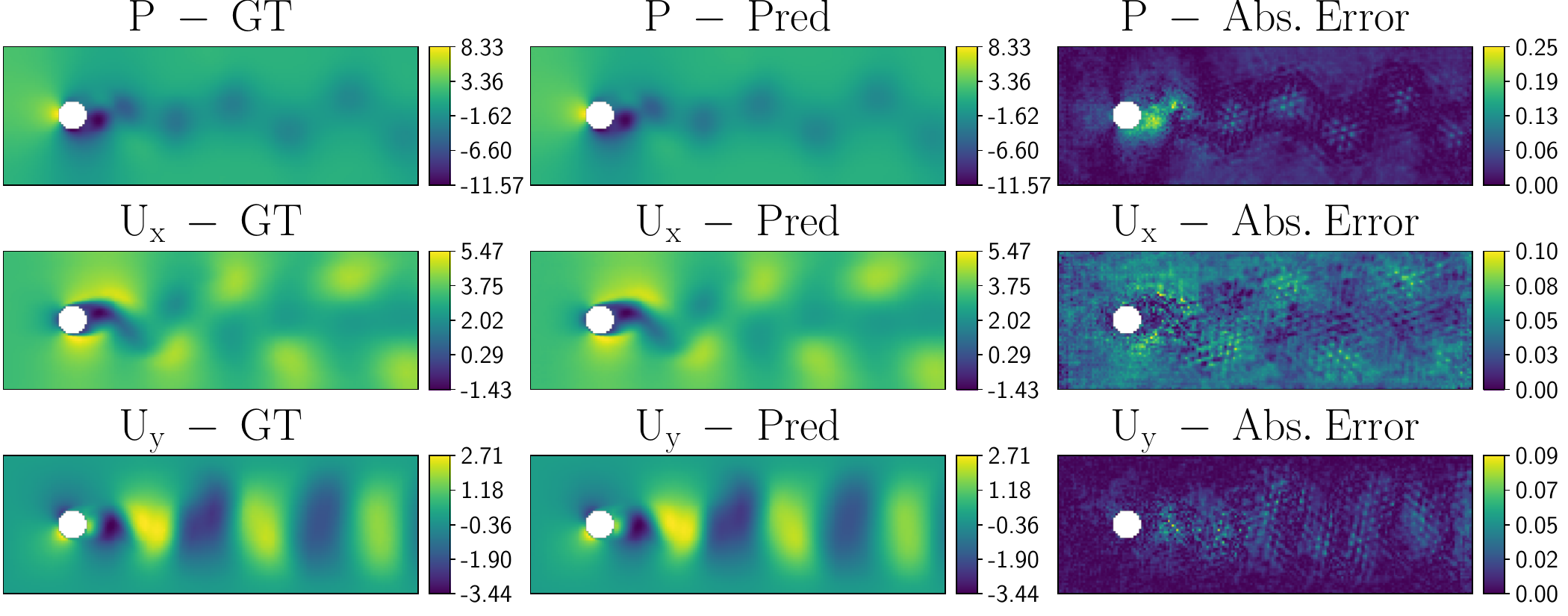}
				\caption{High resolution}
				\label{fig:aae_example_2_newt_hr}
	   		\end{subfigure}
	   \end{subfigure}
	\caption{Results of the prediction made by the Adversarial Autoencoder (AAE). \ref{fig:aae_example_1_newt_lr}: Low resolution Ground Truth (GT), AAE prediction and absolute error for $P$, $U_{x}$ and $U_{y}$. The ground truth is the input of the AAE and it predicts the low resolution fields and the high resolution fields shown in Fig. \ref{fig:aae_example_1_newt_hr}. \ref{fig:aae_example_1_newt_hr}: High resolution GT, AAE prediction and absolute error fields for the same snapshot shown in Fig. \ref{fig:aae_example_1_newt_lr}. \ref{fig:aae_example_2_newt_lr}: Low resolution fields for a different snapshots. \ref{fig:aae_example_2_newt_hr}: High resolution GT, AAE pred. and abs. error fields for the snapshot shown in Fig. \ref{fig:aae_example_2_newt_lr} (cont.).}
	\label{fig:aae_newt_predictions2}
\end{figure}

In order to prove the convenience of our proposed method we compare it to classical resolution augmentation techniques. The AAE is compared with a common technique to augment resolution in the computer vision field: the bicubic interpolation. Comparison between both is shown in Table \ref{tab:aae_vs_bicinterp_newt}. The results of the AAE outperform the bicubic interpolation, while being considerably faster, leading to a $37\times$ speed increment.

\begin{figure}[h]
	\centering
		\begin{subfigure}{0.49\textwidth}
			\includegraphics[width=\textwidth]{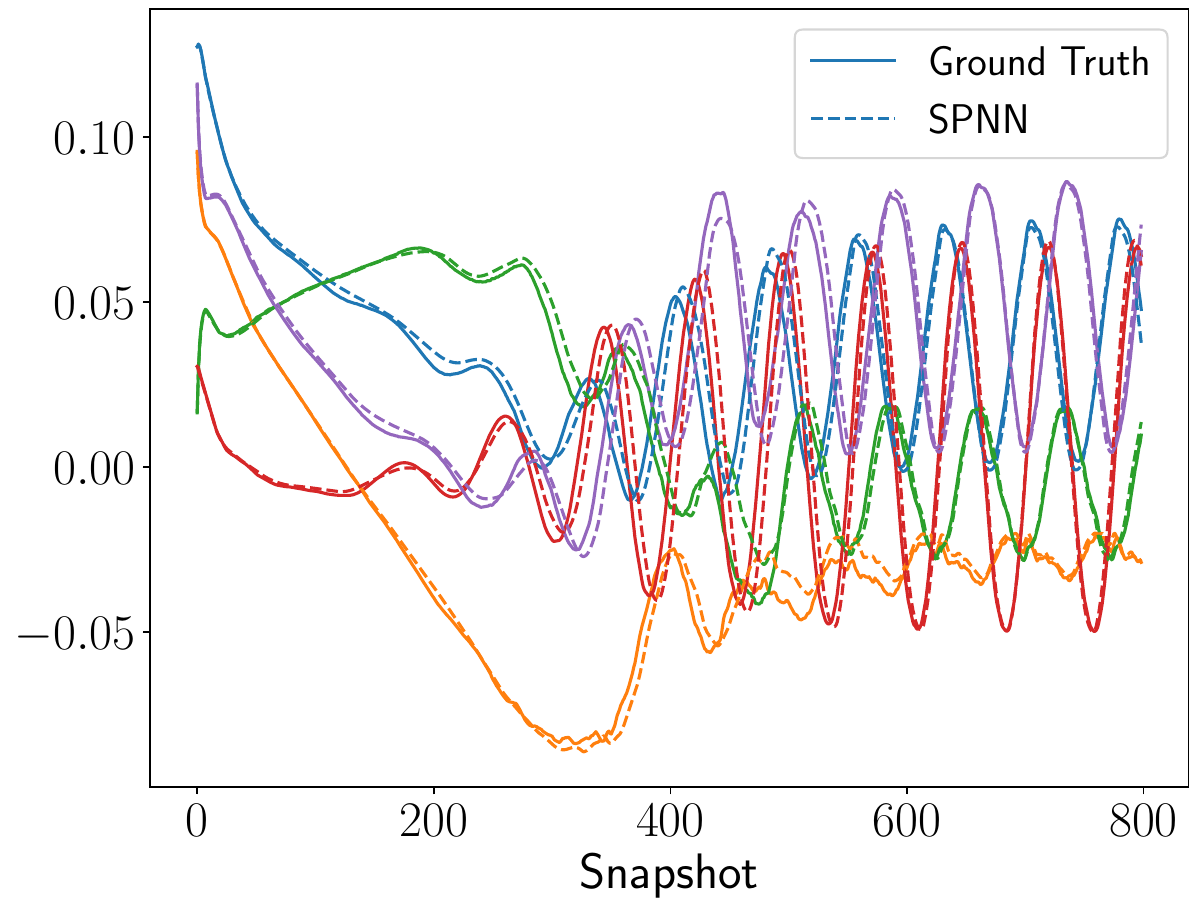}
			\label{fig:spnn_gt_newt_1}
	   \end{subfigure}
	   \begin{subfigure}{0.49\textwidth}
			\includegraphics[width=\textwidth]{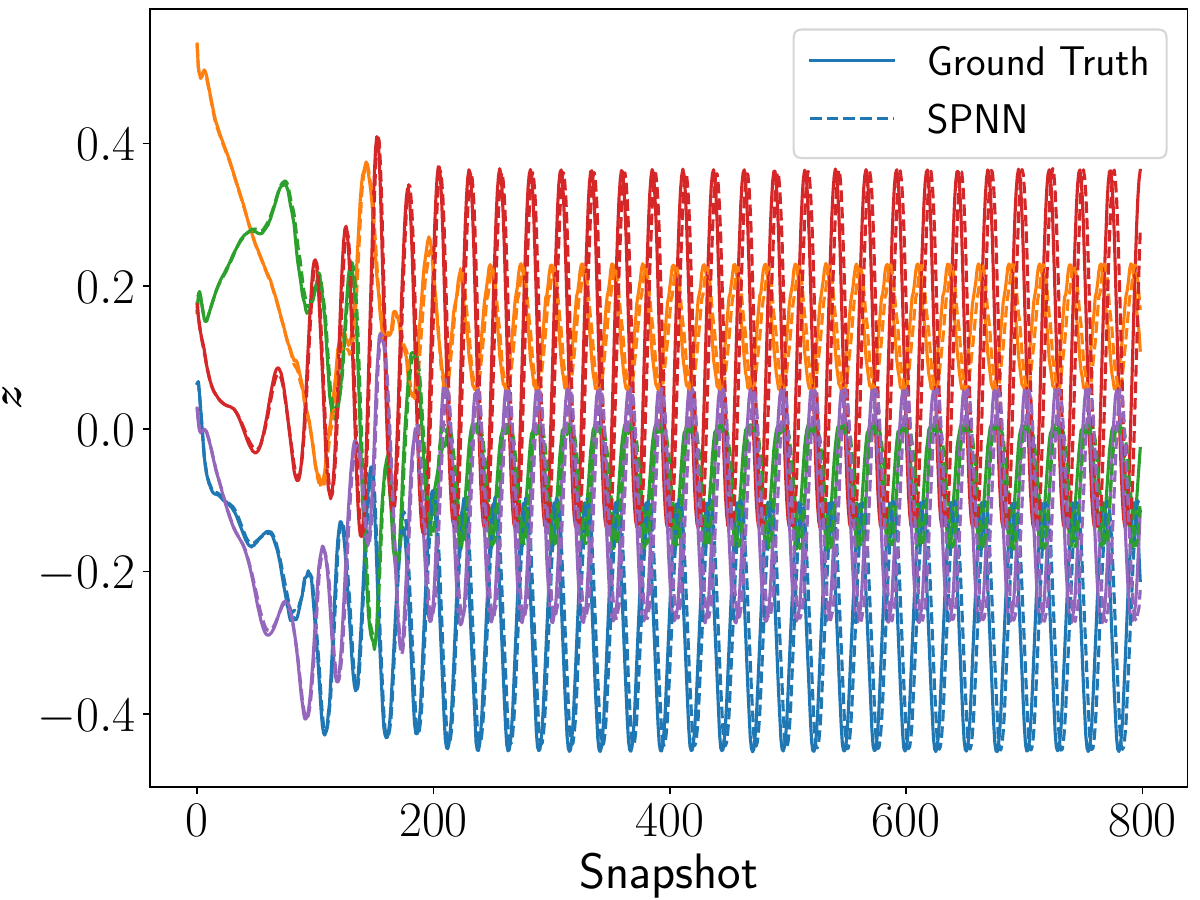}
			\label{fig:spnn_gt_newt_2}
	    \end{subfigure}
	\caption{Results of the SPNN integration with respect to the ground truth (GT) latent variables obtained by the AAE for two different simulations of the newtonian fluid. The dashed line corresponds to the SPNN prediction while the continuous line is the ground truth. To facilitate the identification of ground truth and prediction values, each latent variable is represented by a distinct color. This simplifies the comparison between the ground truth and the SPNN prediction. Left: simulation at $\mathrm{U_{in}=1.4}$. Right: simulation at $\mathrm{U_{in}=3.4}$.}
	\label{fig:spnn_gt_newt}
\end{figure}

\begin{table}[h!]
	\begin{center}
		\captionsetup{skip=10pt}
		\caption{Comparison between the proposed method and bicubic interpolation. For every variable, mean relative error is shown, and the reconstruction time for the whole dataset is computed.}
			\begin{tabular}{| c | c | c |}
				\hline
				 & AAE & Bicubic interpolation \\ \hline
				$P$ (-) & 0.0247 &  0.0621 \\
				$U_{x}$ (-) & 0.0105 & 0.0285 \\
				$U_{y}$ (-) & 0.0196 & 0.0524 \\ 
				Time ($s$) & 711.05 & 26303.31 \\ \hline
			\end{tabular}
		\label{tab:aae_vs_bicinterp_newt}
	\end{center}
\end{table}
Fig. \ref{fig:spnn_gt_newt} shows the comparison between the AAE latent variables and the rollout prediction made by the SPNN, for the case with input velocity $u=3.4$---the worst case scenario among all considered---. The SPNN is able to integrate the latent variables in the reduced space in good agreement with the original AAE encoding, considered as the ground truth for the SPNN.

\begin{figure}[h]
	\centering
		\begin{subfigure}{0.49\textwidth}
			\includegraphics[width=\textwidth]{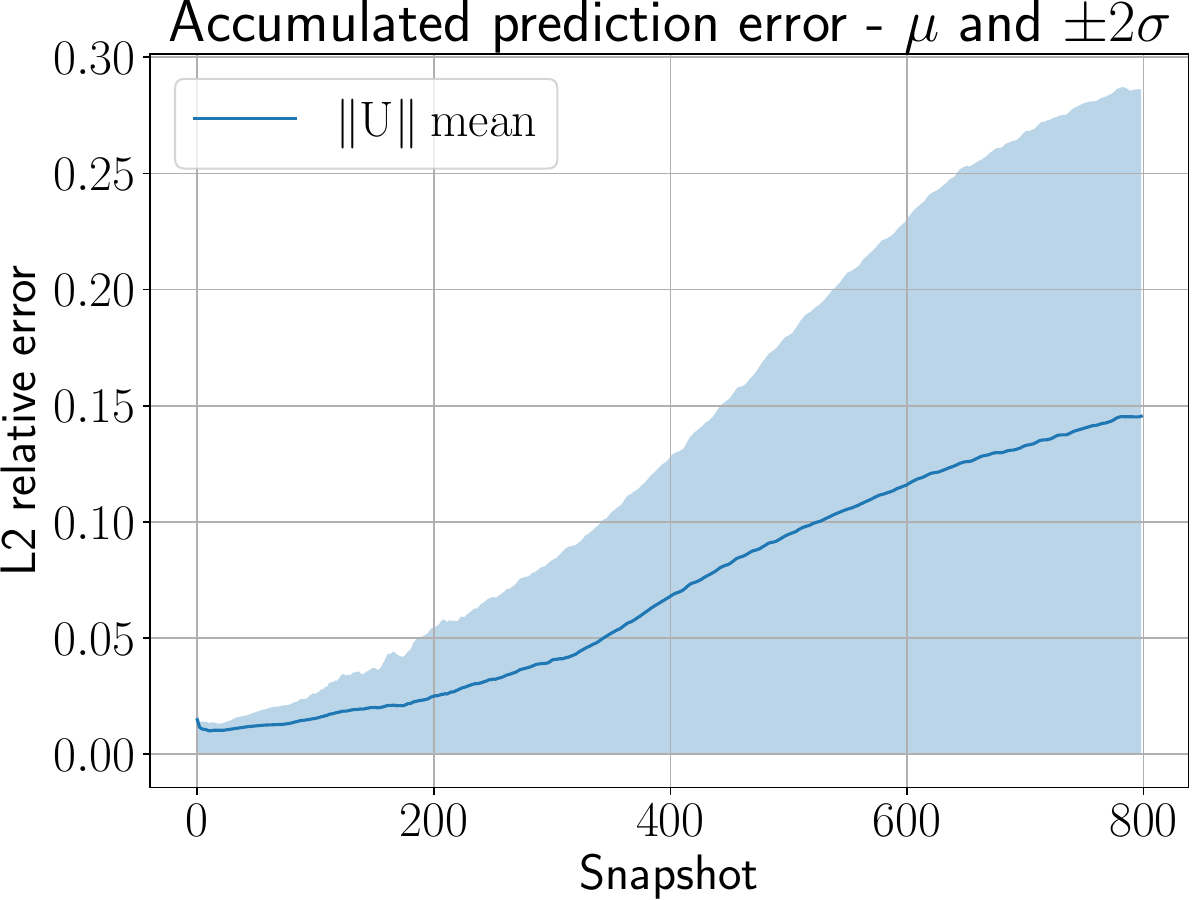}
			\caption{SPNN}
			\label{fig:spnn_newt_evolution}
	   \end{subfigure}
	   \begin{subfigure}{0.49\textwidth}
			\includegraphics[width=\textwidth]{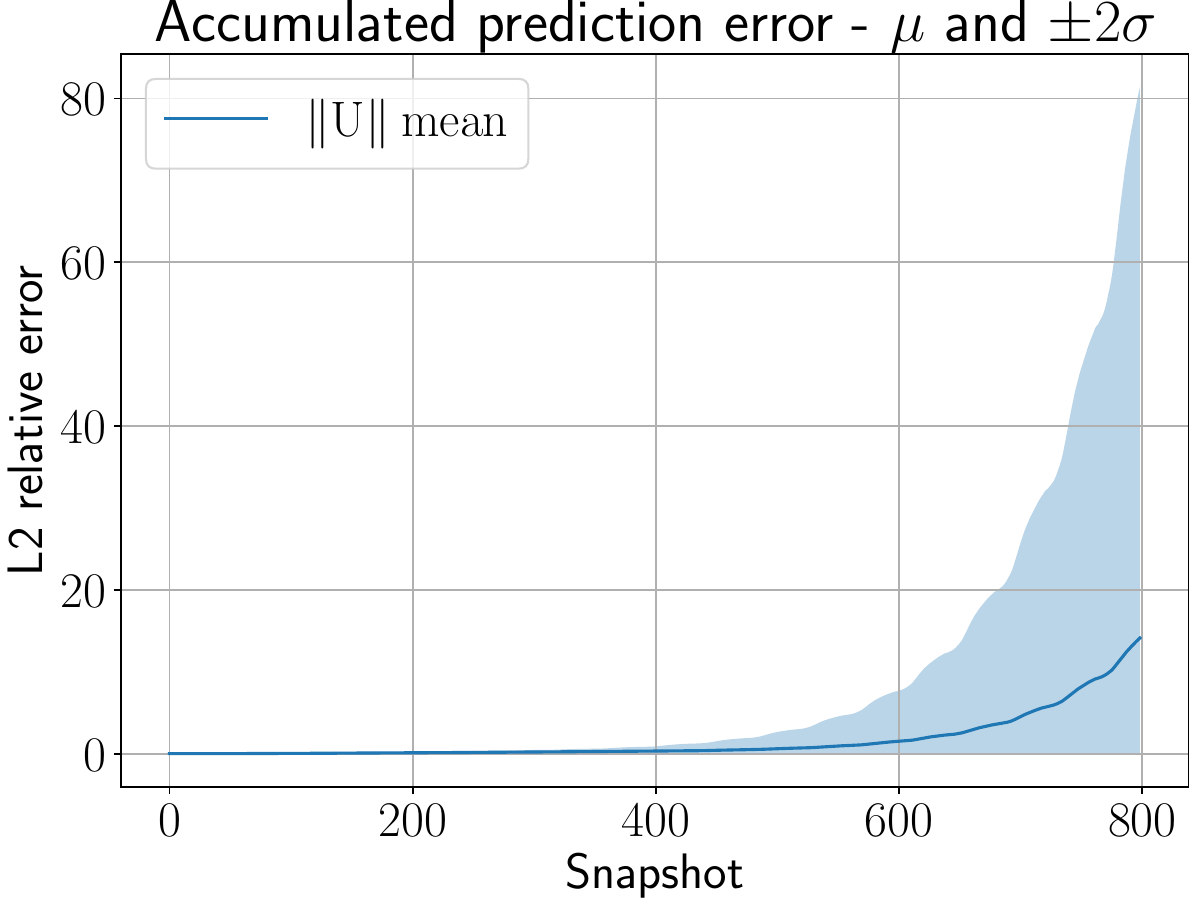}
			\caption{Black box}
			\label{fig:blackbox_newt_evolution}
	    \end{subfigure}
	\caption{Evolution of the relative error during the whole integration of the velocity for all the simulations. Even if the error of the SPNN increases as a consequence of the explicit Euler integration scheme employed, the black box approach fails during the rollout and diverges from the solution. The SPNN error accumulates during the rollout prediction but its able to converge to a meaningful solution.}
	\label{fig:blackbox_results_newt}
\end{figure}

The SPNN is also compared with a black box (BB) neural network. The black box neural network predicts the increment of the latent variables  $\mbox{\boldmath$z$}$ and uses a forward Euler integration scheme to obtain the next snapshot of the simulation. The black box is trained using the same hyperparameters as the SPNN, except for the output size, which is the same dimension as the input, for this example $N_{\text{in}}^{\text{BB}} = N_{\text{out}}^{\text{BB}} = 5$. Results are compared in Fig. \ref{fig:blackbox_results_newt}, which shows the velocity accumulated error mean and standard deviation with a confidence interval of 95\% for every simulation at each snapshot. The SPNN (Fig. \ref{fig:spnn_newt_evolution}) error raises as the prediction advances, which was expected, as the forward Euler is a first order integration scheme, but the network is able to converge to the solution. Meanwhile, the black box neural network (Fig. \ref{fig:blackbox_newt_evolution}) is not able to integrate the predicted trajectory and diverges from the ground truth, proving that the thermodynamic bias guides the network to converge to a meaningful solution.

The ground truth simulations were performed on a MacBook Pro M1 Pro. Each simulation took around 20 minutes to complete. The AAE and the SPNN were trained using the Pytorch framework. The computer used to train both networks was a Linux-based machine equipped with a Intel i9-13900K CPU and a NVIDIA RTX 4090 GPU. The AAE training time was approximately 4 hours, while the SPNN took around 20 minutes to train. While working on inference, the prediction for the latent variables can be obtained in 1-2 seconds in a MacBook Pro M1 Pro, while rendering the video for the complete simulation takes around 15 seconds, achieving a considerable speedup when compared with the computational cost of running the high-fidelity simulation.
\subsection{Example 2: Flow past a cylinder of a non-Newtonian fluid}
\label{subsec:example2}

\subsubsection{Database generation}
\label{subsubsec:data2}

The second example is generated by using the same geometry than in the previous example, but the fluid is replaced by a non-Newtonian fluid. As in the previous case, the flow conditions are obtained by varying the initial velocity of the flow, which results in different Reynolds numbers. The state variables for the non-Newtonian flow past a cylinder are the velocity, shear rate and pressure fields, although good prediction results can be achieved by using only the pressure and velocity fields obtained from the solver:
\begin{equation} \label{eq:state_var_non_newt2}
	\mathcal{S} = \lbrace \mbox{\boldmath$x$} = \left( \mbox{\boldmath$u$}, \dot{\gamma},P \right) \in \mathbb{R}^{2} \times \mathbb{R} \times \mathbb{R} \rbrace.
\end{equation}

Ground truth simulations are obtained by solving the 2D Navier-Stokes equations using OpenFOAM \cite{OPENFOAM}. A no-slip condition is applied at the wall of the cylinder. In this example, a non-Newtonian fluid behaviour is applied using the Herschel-Bulkey model in OpenFOAM, defined by the following parameters: $\rho = 1$, $\nu_{0} = 0.00125$, $\tau_{0} = 0.00125$, $k = 0.015625$, $n = 1.88$. The freestream velocity is contained within the interval \( \mbox{\boldmath$u$} \in \left[ 1.0, 2.0 \right] \), with speed increments of \( 0.1 \), which results in a total of $N_{\text{sim}} = 11$ cases. Each simulation is discretized in $\tt n_{\text{snap}}= 600$ time increments of $\Delta t = 0.005$.

\begin{figure}[h]
	\centering
		\begin{subfigure}{0.49\textwidth}
			\includegraphics[width=\textwidth]{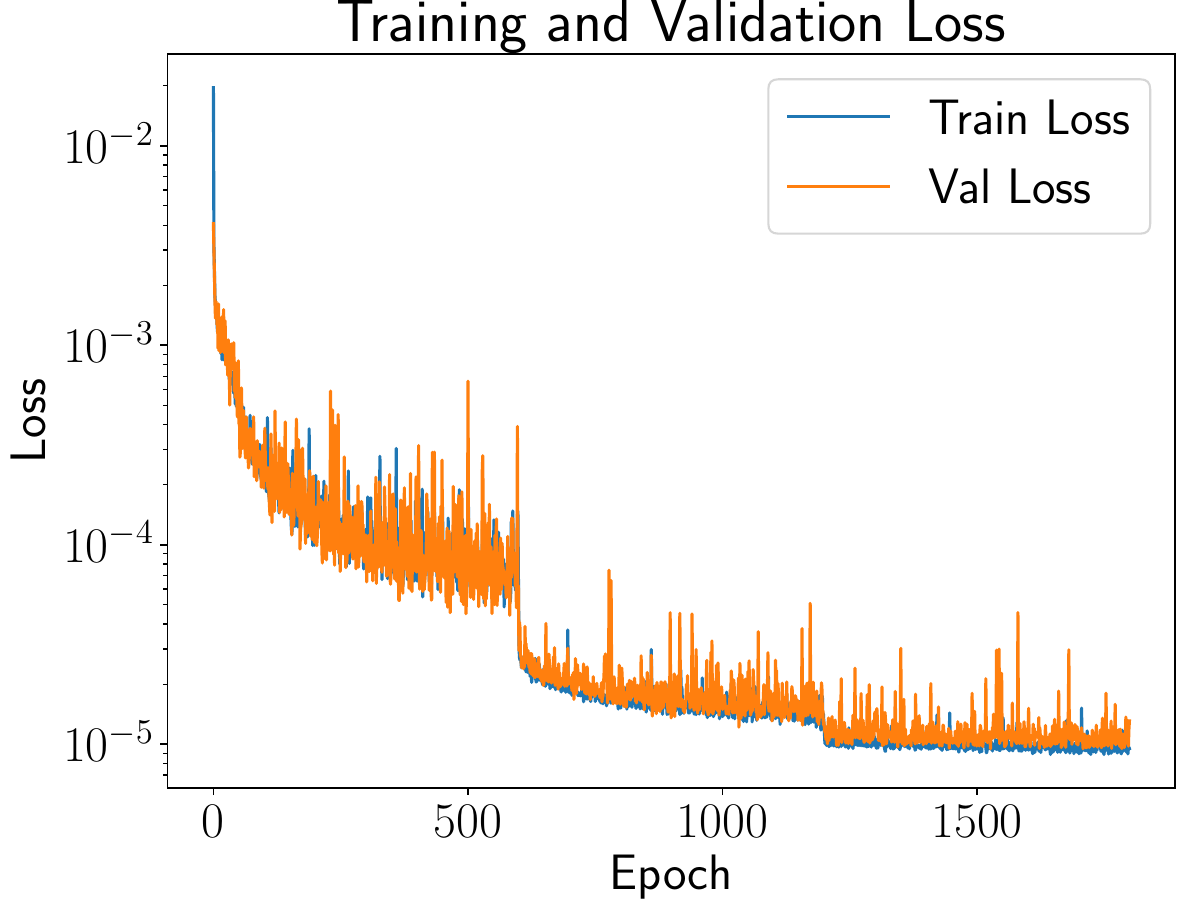}
            \caption{AAE Loss}
			\label{fig:aae_loss_non_newt}
	   \end{subfigure}
	   \begin{subfigure}{0.49\textwidth}
			\includegraphics[width=\textwidth]{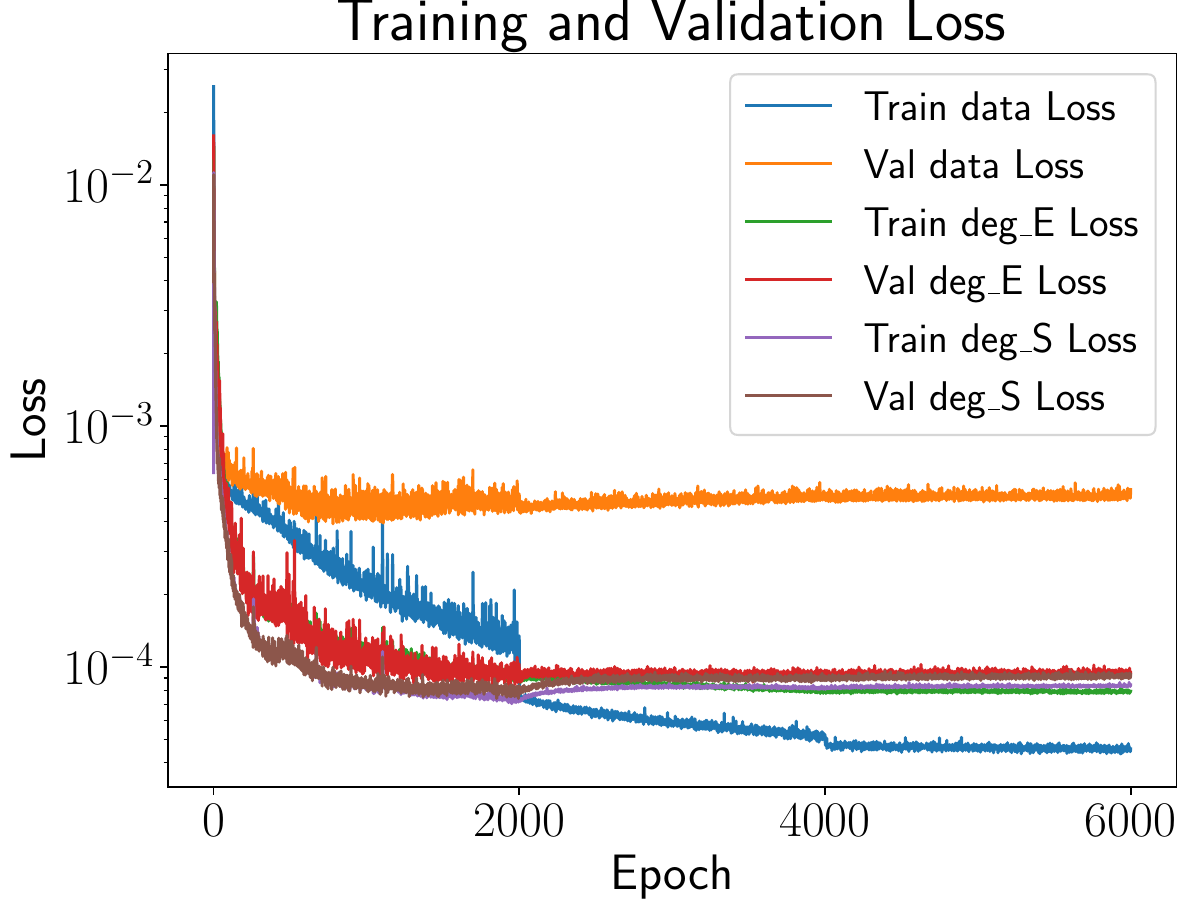}
            \caption{SPNN Loss}
			\label{fig:spnn_loss_non_newt}
	    \end{subfigure}
	\caption{Adversarial Autoencoder (left) and the SPNN (right) training and validation loss curves for the non-Newtonian example.}
	\label{fig:loss_non_newt}
\end{figure}

The input of the autoencoder are the low resolution velocity and pressure fields, with size $3 \times 16 \times 48$ , while the output are the velocitiy and pressure fields at the original resolution and a higher one, with sizes  $3 \times 16 \times 48$ and $3 \times 64 \times 192$. Both the encoder and decoder use convolutional layers with $ N_{\text{ch}} = 64$ channels an a kernel size of $k = 3$, following a ResNet-like structure \cite{RESNET}. The number of latent variables at the bottleneck is set to $d = 6$. The activation function used is the leaky-ReLU with a negative slope of $0.1$, except for the last layer of both the encoder and decoder, where linear activations are used. The adversarial hyperparameter weight is set to $\lambda_{\text{adv}}^{\text{AAE}} = 10^{-3}$. The optimizer used is Adam \cite{ADAM} with a learning rate set to $l_{\text{r}}^{\text{AAE}} = 10^{-4}$ with decreasing order of magnitude on epochs 600 and 1200, a weight decay set to $w_{\text{d}}^{\text{AAE}} = 10^{-6}$, and a total number of $N_{\text{epochs}} = 1800$ epochs. 
Latent variables obtained at the bottleneck are then used as input variables for the structure preserving neural network. Fig. \ref{fig:aae_loss_non_newt} shows the training and validation loss for the adversarial autoencoder.

The SPNN input size coincides with the AAE latent dimension, $N_{\text{in}}^{\text{SPNN}} = d = 6$, while the output size is $N_{\text{out}}^{\text{SPNN}} = d \cdot \left( d + 2 \right) = 48$. The number of hidden layers of the SPNN is $N_{\text{hl}}^{\text{SPNN}}=5$ with 120 neurons each one, Leaky-ReLU activations and linear for the last layer. The data weight hyperparameter is set to $\lambda_{\text{data}}^{\text{SPNN}} = 10^{2}$. The SPNN is trained for $N_{epochs} = 6000$ epochs using the Adam optimizer, and a batch size of $B_{\text{size}} = 128$. The learning rate is set to  $l_{\text{r}}^{\text{SPNN}} = 10{-3}$, decreasing one order of magnitude on epoch 2000 and 4000. The weight decay is set to $w_{\text{d}}^{\text{SPNN}} = 10^{-4}$ and noise variance is added to the train set, $\sigma_{\text{noise}}^{2} = 10^{-5}$. Fig. \ref{fig:spnn_loss_non_newt} shows the training and validation loss for the SPNN.

\subsubsection{Results}
\label{subsubsec:results_ex2}

\begin{figure}[h]
	\centering
		\begin{subfigure}{0.49\textwidth}
			\includegraphics[width=\textwidth]{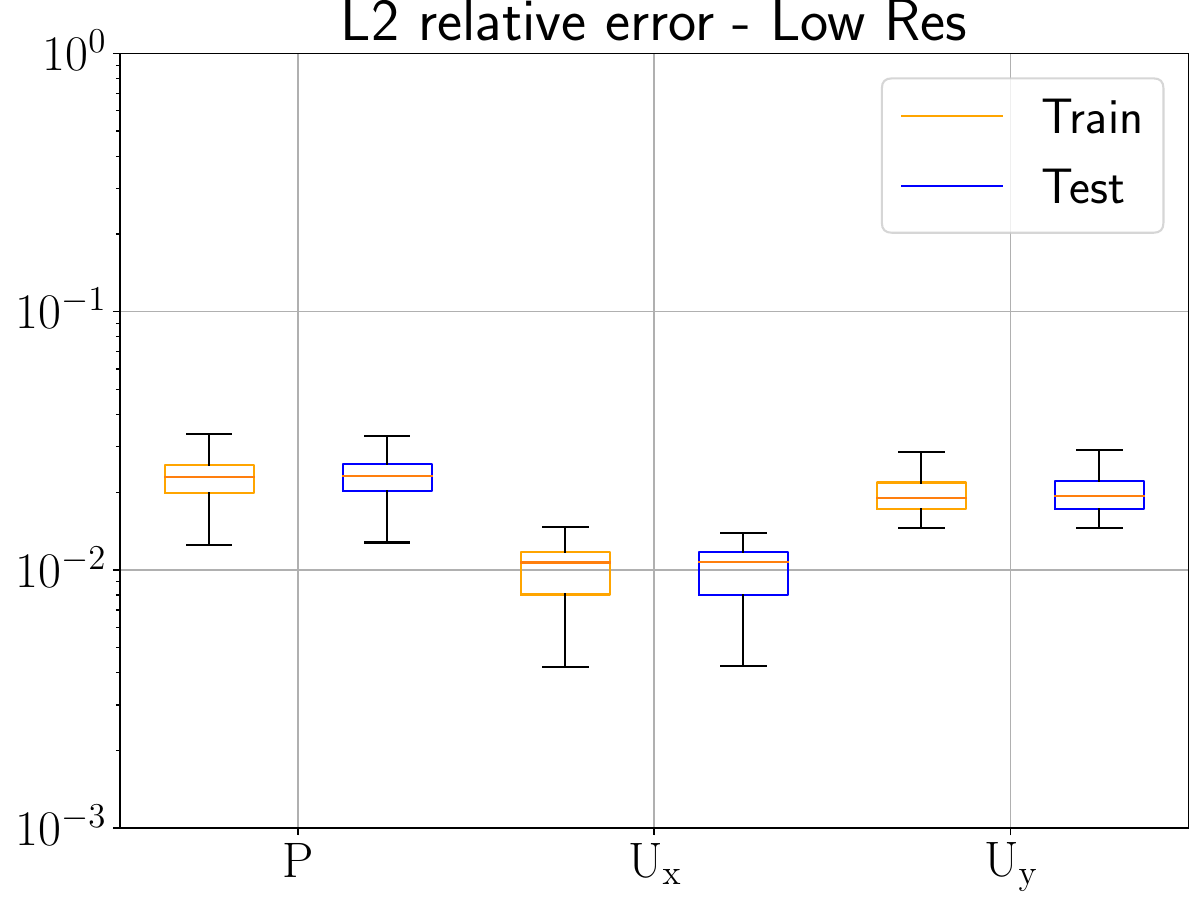}
			\label{fig:relative_error_aae_non_newt_lr}
	   \end{subfigure}
	   \begin{subfigure}{0.49\textwidth}
			\includegraphics[width=\textwidth]{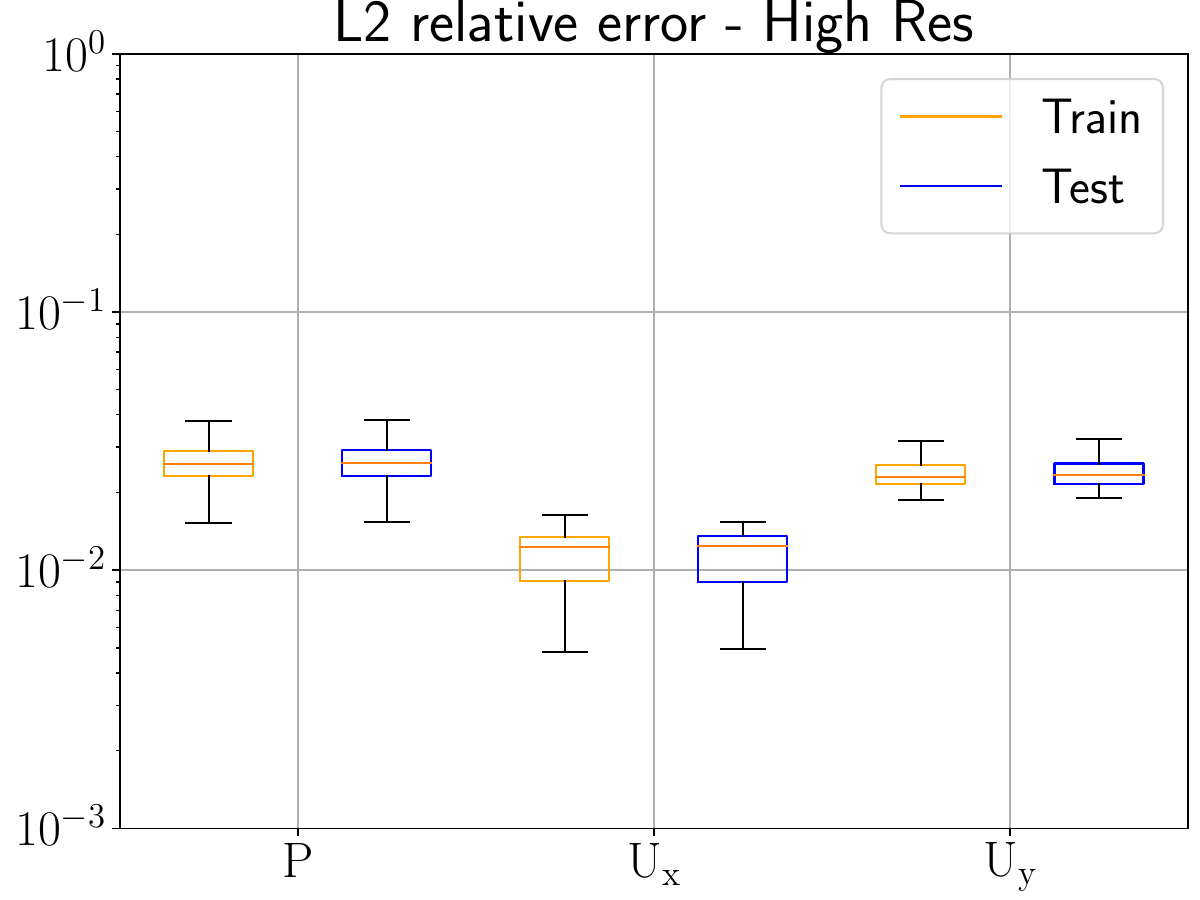}
			\label{fig:relative_error_aae_non_newt_hr}
	    \end{subfigure}
	\caption{Box plots for the relative L2 error of the autoencoder for all the snapshots of the non-newtonian fluid for both train and test cases, in low (left) and high (right) resolution.}
	\label{fig:aae_non_newt_relative_error}
\end{figure}

Fig. \ref{fig:aae_non_newt_predictions} shows the prediction results for the pressure and velocity fields obtained by the autoencoder for both, low and high resolution, as well as the absolute error for each field. The  low and high resolution fields reconstructed by the autoencoder show good agreement with the ground truth fields obtained from the CFD simulation. A box plot containing the error of the state variables for the train and test cases is shown in Fig. \ref{fig:aae_non_newt_relative_error}, achieving a mean error lower than 3\% for the pressure and velocity fields in both low and high resolution.
\begin{figure}[h]
	\centering
		\begin{subfigure}{\textwidth}
			\begin{subfigure}{\textwidth}
				\includegraphics[width=\textwidth]{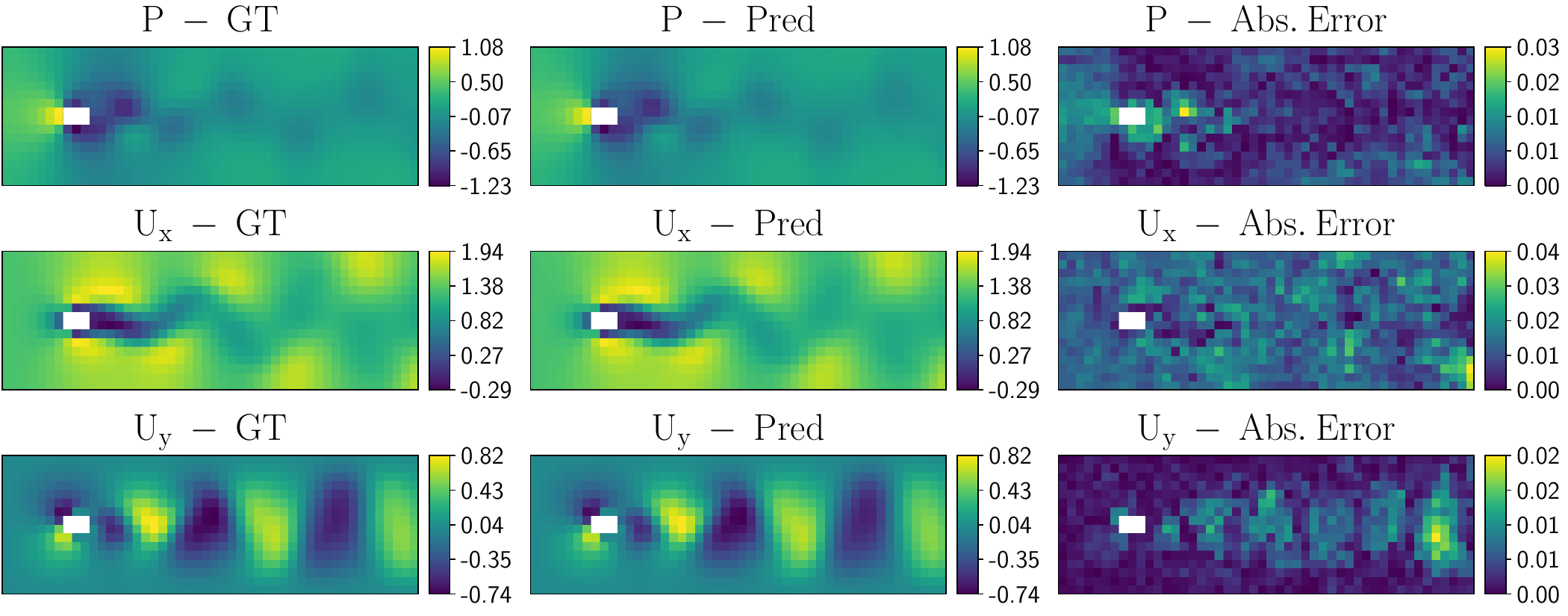}
				\caption{Low resolution}
				\label{fig:aae_example_1_non_newt_lr}
	   		\end{subfigure}
	   		\vfill
	   		\begin{subfigure}{\textwidth}
				\includegraphics[width=\textwidth]{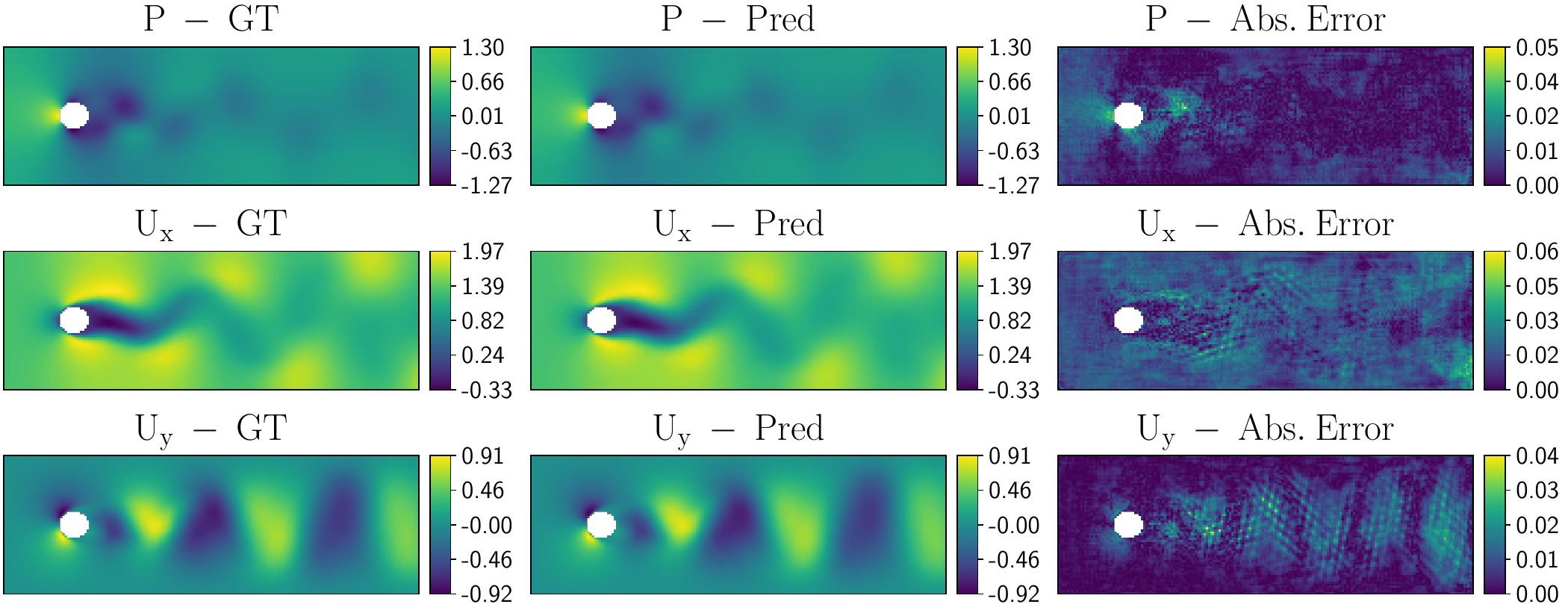}
				\caption{High resolution}
				\label{fig:aae_example_1_non_newt_hr}
	   		\end{subfigure}
	   \end{subfigure}
	\caption{Results of the AAE for two different snapshots. \ref{fig:aae_example_1_non_newt_lr}: Low resolution Ground Truth (GT), AAE prediction and absolute error for $P$, $U_{x}$ and $U_{y}$ fields for the first snapshot. \ref{fig:aae_example_1_non_newt_hr}: High resolution GT, AAE prediction and absolute error fields for the same snapshot shown in Fig. \ref{fig:aae_example_1_non_newt_lr}. \ref{fig:aae_example_2_non_newt_lr}: Low resolution fields for the second snapshot. \ref{fig:aae_example_2_non_newt_hr}: High resolution fields for the snapshot shown in Fig. \ref{fig:aae_example_2_non_newt_lr}.}
	\label{fig:aae_non_newt_predictions}
\end{figure}
\begin{figure}[h]\ContinuedFloat
	\centering
		\begin{subfigure}{\textwidth}
			\begin{subfigure}{\textwidth}
				\includegraphics[width=\textwidth]{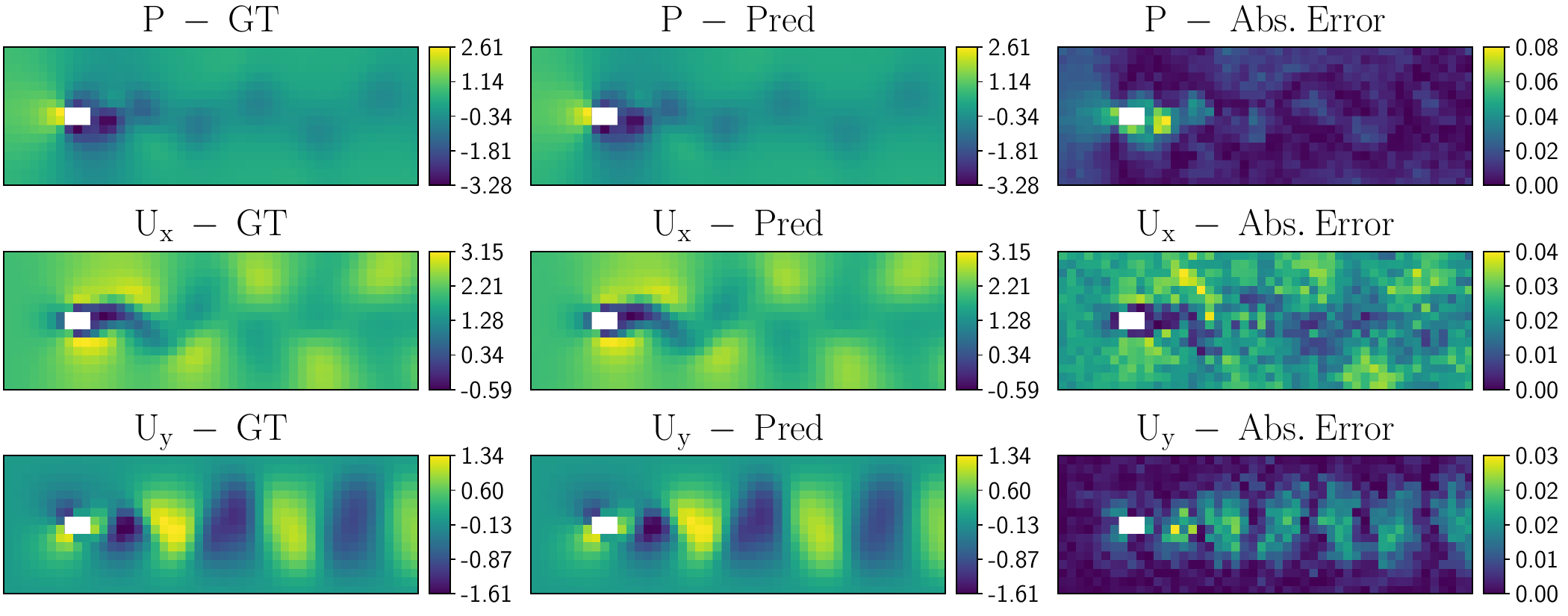}
				\caption{Low resolution}
				\label{fig:aae_example_2_non_newt_lr}
	   		\end{subfigure}
	   		\vfill
	   		\begin{subfigure}{\textwidth}
				\includegraphics[width=\textwidth]{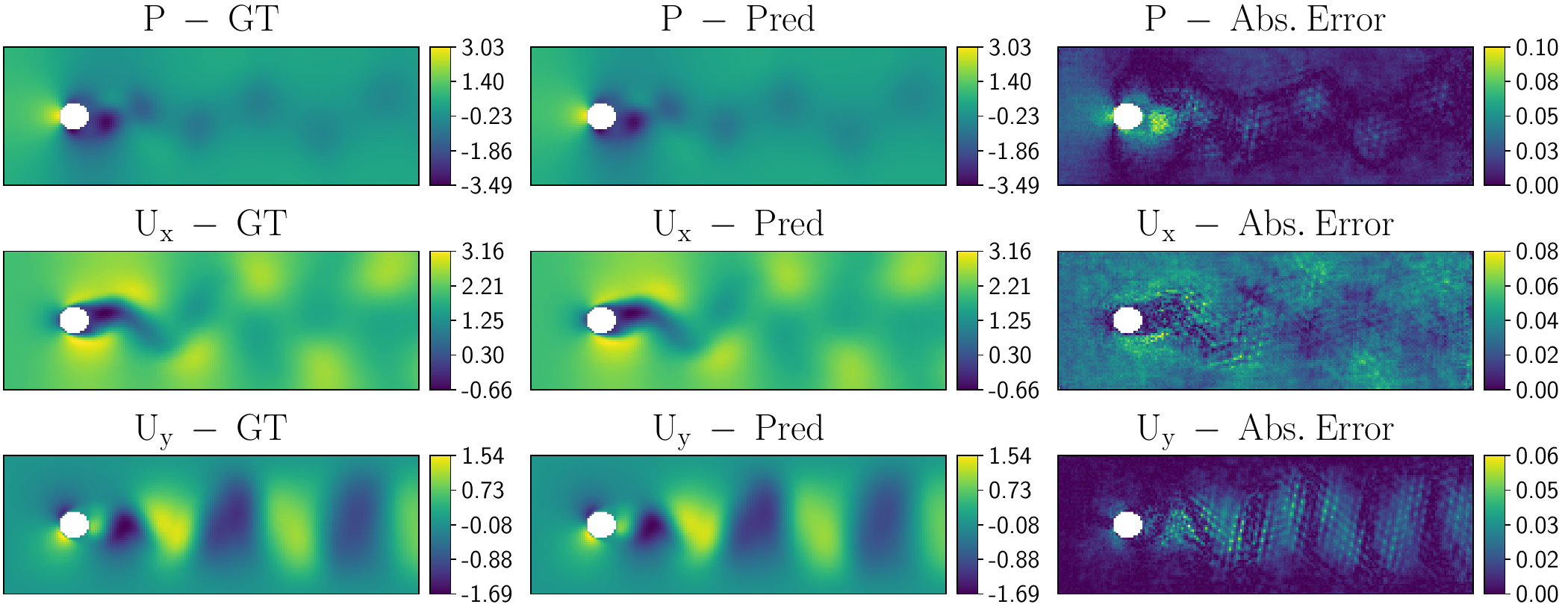}
				\caption{High resolution}
				\label{fig:aae_example_2_non_newt_hr}
	   		\end{subfigure}
	   \end{subfigure}
	 \caption{Results of the AAE for two different snapshots. \ref{fig:aae_example_1_non_newt_lr}: Low resolution Ground Truth (GT), AAE prediction and absolute error for $P$, $U_{x}$ and $U_{y}$ fields for the first snapshot. \ref{fig:aae_example_1_non_newt_hr}: High resolution GT, AAE prediction and absolute error fields for the same snapshot shown in Fig. \ref{fig:aae_example_1_non_newt_lr}. \ref{fig:aae_example_2_non_newt_lr}: Low resolution fields for the second snapshot. \ref{fig:aae_example_2_non_newt_hr}: High resolution fields for the snapshot shown in Fig. \ref{fig:aae_example_2_non_newt_lr} (cont.).}
	\label{fig:aae_non_newt_predictions2}
\end{figure}

\begin{figure}[h]
	\centering
		\begin{subfigure}{0.49\textwidth}
			\includegraphics[width=\textwidth]{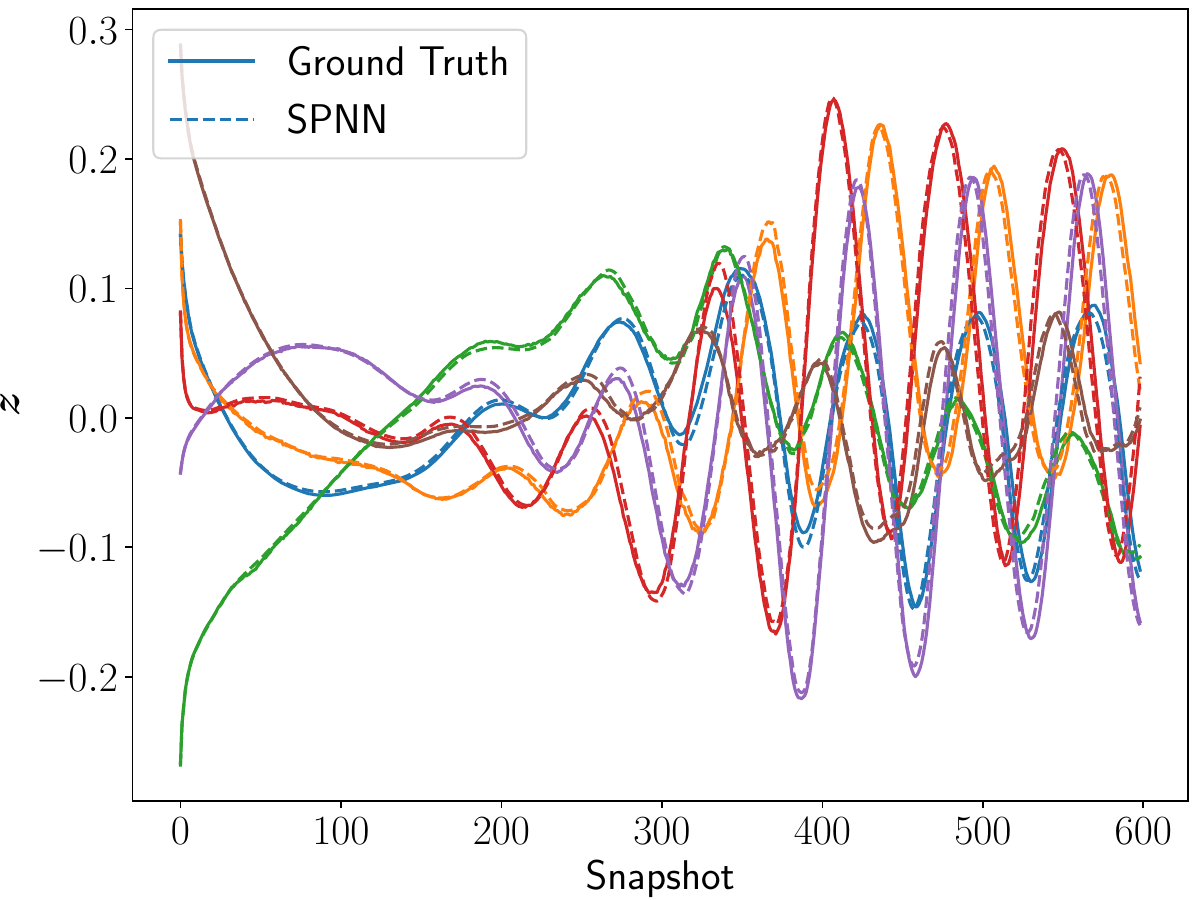}
			\label{fig:spnn_gt_non_newt_1}
	   \end{subfigure}
	   \begin{subfigure}{0.49\textwidth}
			\includegraphics[width=\textwidth]{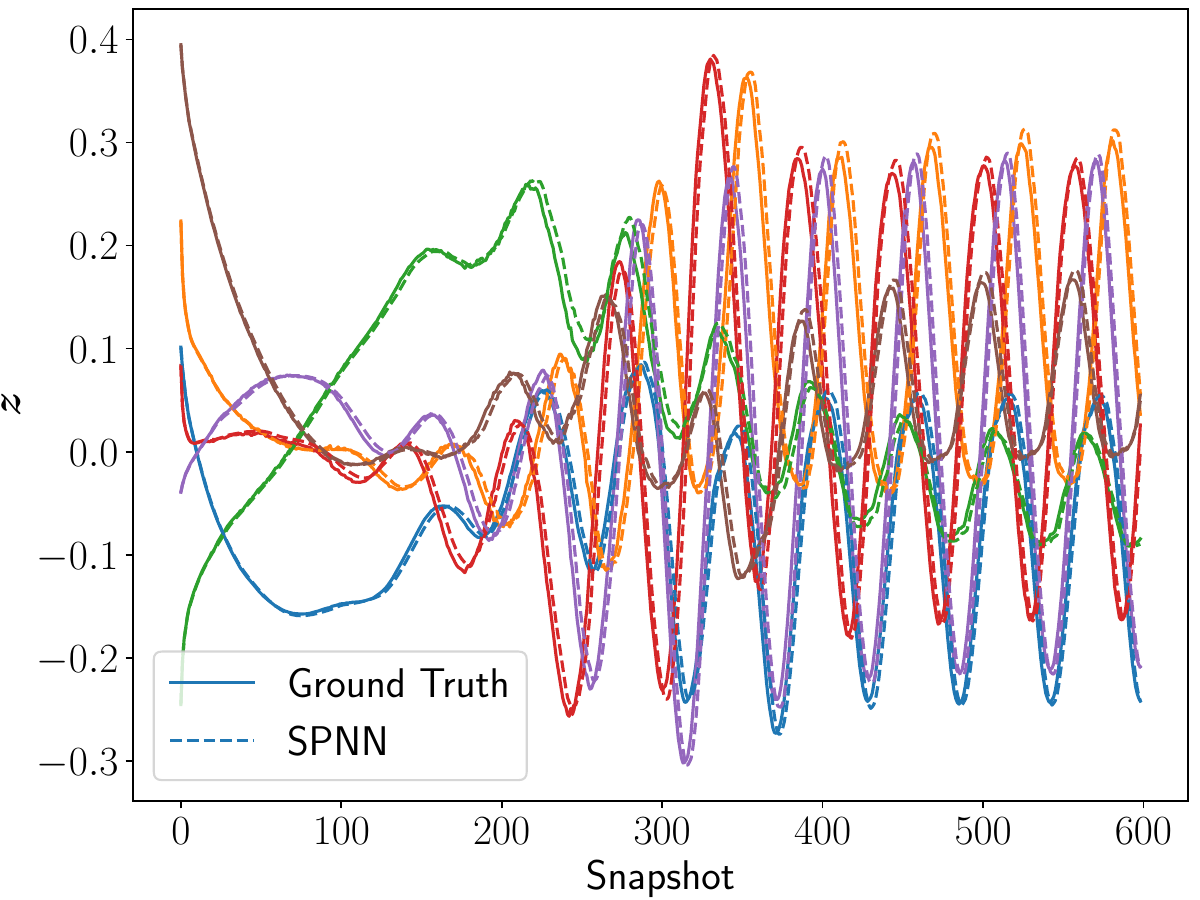}
			\label{fig:spnn_gt_non_newt_2}
	    \end{subfigure}
	\caption{Results of the SPNN integration with respect to the latent variables obtained by the AAE (ground truth) for two different simulations for the non-Newtonian fluid. The dashed line corresponds to the SPNN prediction while the continuous line is the ground truth. Each latent variable is represented in a different color to simplify the identification of the ground truth and its corresponding predicted value. Left: simulation at $\mathrm{U_{in}=1.4}$. Right: simulation at $\mathrm{U_{in}=1.7}$}
	\label{fig:spnn_gt_non_newt}
\end{figure}

As in the previous case, the autoencoder has been compared with the bicubic interpolation technique. Comparison is shown in Table \ref{tab:aae_vs_bicinterp_non_newt}. As expected, considering the results obtained for the newtonian fluid case, the AAE clearly outperforms the bocubic interpolation, specially if the time difference between both methods is considered, with the AAE being almost 45 times faster than the bicubic interpolation.
\begin{table}[h]
	\begin{center}
		\captionsetup{skip=10pt}
		\caption{Comparison between the AAE and bicubic interpolation. For every variable, relative error is shown, and time reconstruction time for all dataset is computed.}
		\begin{tabular}{| c | c | c |}
			\hline
			 & AAE & Bicubic interpolation \\ \hline
			$P$ (-) & 0.0257 &  0.0557 \\
			$U_{x}$ (-) & 0.0113 & 0.0228 \\
			$U_{y}$ (-) & 0.0240 & 0.0639 \\ 
			Time ($s$) & 197.57 & 8818.12 \\ \hline
		\end{tabular}
		\label{tab:aae_vs_bicinterp_non_newt}
	\end{center}
\end{table}
Fig. \ref{fig:spnn_gt_non_newt} shows the comparison between the ground truth latent variables, the ones obtained by the AAE and the prediction made by the SPNN. As in the previous case, the SPNN is able to integrate the latent variables in the reduced space successfully with respect to the original AAE encoding.

\begin{figure}[h]
	\centering
		\begin{subfigure}{0.49\textwidth}
			\includegraphics[width=\textwidth]{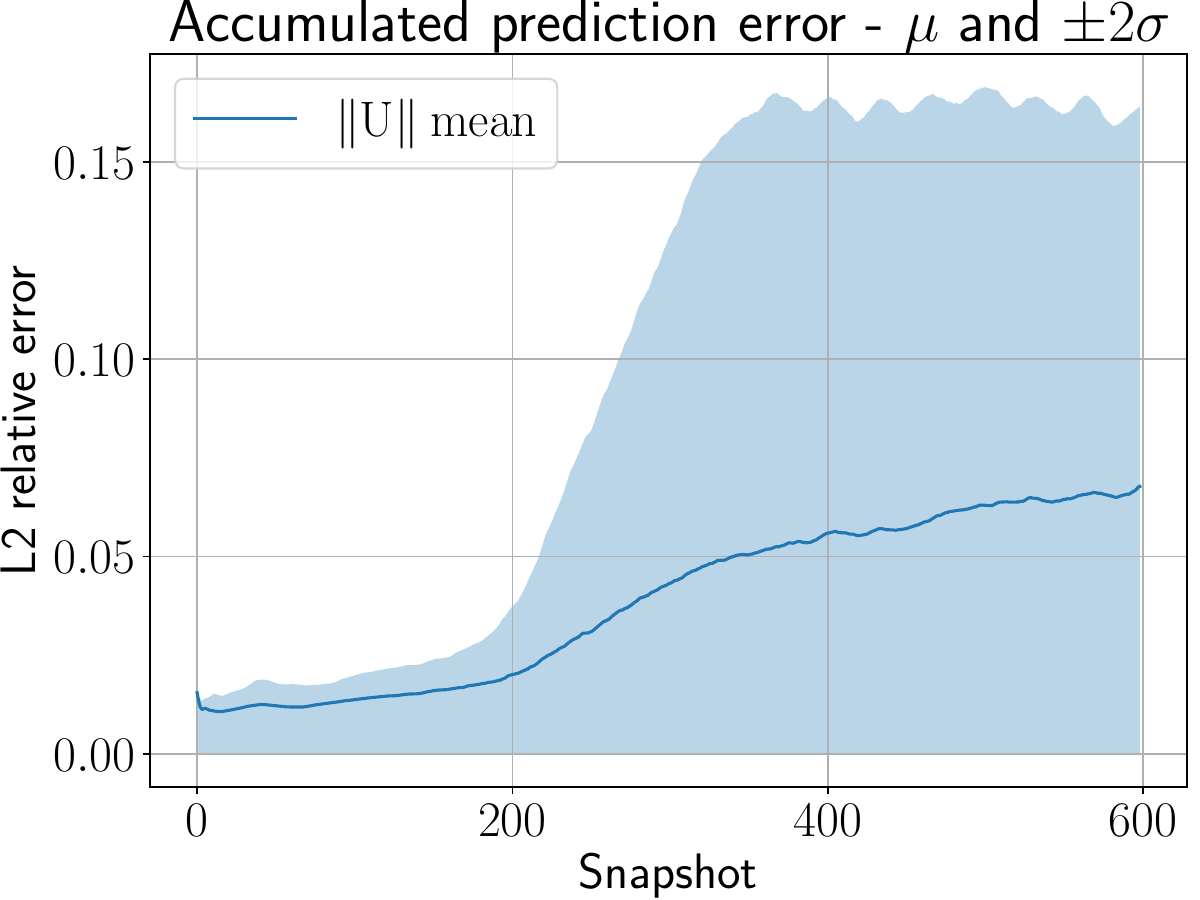}
			\caption{SPNN}
			\label{fig:spnn_non_newt_evolution}
	   \end{subfigure}
	   \begin{subfigure}{0.49\textwidth}
			\includegraphics[width=\textwidth]{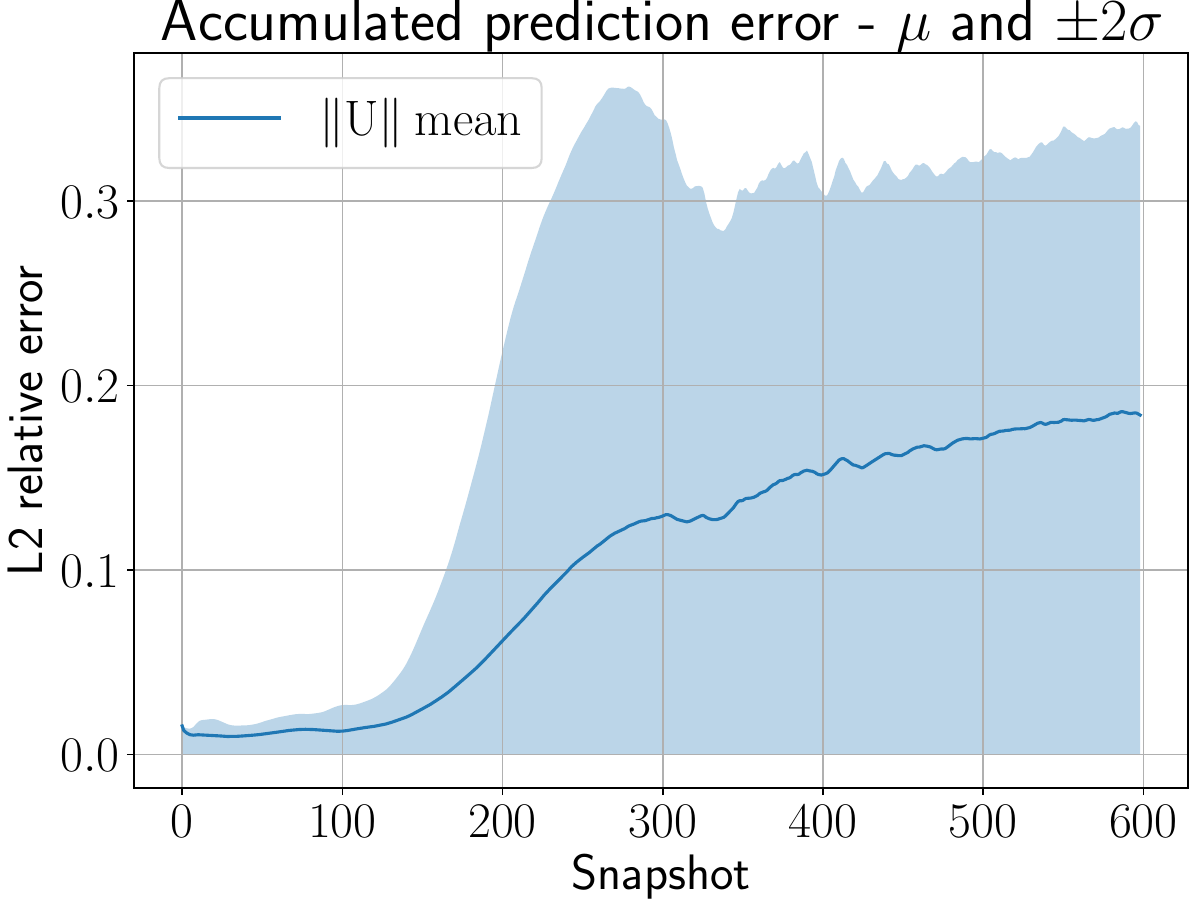}
			\caption{Black box}
			\label{fig:blackbox_non_newt_evolution}
	    \end{subfigure}
	\caption{Evolution of the relative error during the whole integration of the velocity for all the simulations. The SPNN error remains lower than the black box neural network during all the prediction rollout, proving that the physical bias helps to reach the correct solution.}
	\label{fig:blackbox_results_non_newt}
\end{figure}

As previously, the SPNN is compared with a black box neural network, which is trained using the same hyperparameters as the SPNN, except for the outputs size, for this example $N_{\text{in}}^{\text{BB}} = N_{\text{out}}^{\text{BB}} = 6$, the same as the bottleneck of the AAE and the input of the SPNN.  Results are compared in Fig. \ref{fig:blackbox_results_non_newt}, which shows the velocity accumulated error mean and standard deviation with a confidence interval of 95\% for every simulation at each snapshot. The SPNN error (Fig. \ref{fig:spnn_non_newt_evolution}) remains lower than the black box neural network (Fig. \ref{fig:blackbox_non_newt_evolution}), proving that the thermodynamic bias helps the network to converge to the correct solution.

The ground truth simulations were performed on a MacBook Pro M1 Pro, with each simulation taking around 30 minutes to complete. Both, the AAE and the SPNN were trained on a Linux-based machine equipped with a Intel i9-13900K CPU and a NVIDIA RTX 4090 GPU using the Pytorch framework. Training the AAE took 2.5 hours in the NVIDIA GPU, while the SPNN training time was around 20 minutes. Regarding the inference time, each latent variables prediction can be obtained in less than 1 second on a MacBook Pro M1 Pro, although rendering the video with the complete prediction takes about 10 seconds, which is considerably faster than that computing the high-fidelity model.

\section{Conclusions}
\label{sec:conclusions}

In this work we have presented a new methodology to increase the spatial resolution of predictions obtained by learned simulators, while ensuring a thermodynamics-aware prediction, satisfying the basic principles of thermodynamics. The proposed AAE architecture is able to encode the information to a reduced-order space and to produce high-resolution output fields from low resolution input thanks to its generative capabilities. The AAE has been compared to a classical resolution augmentation technique: the bicubic interpolation. Not only the AAE outperforms the bicubic interpolation, but it is also considerably faster, making it feasible for quasi-real-time or even real-time applications. Additionally, AAE resolution augmentation technique can be applied to a wider range of geometries than bicubic interpolation. The structure-preserving neural network is able to estimate the evolution of the encoded variables in the reduced space and then the decoder re-projects the SPNN prediction to the original and higher resolution spaces.
The SPNN is compared to a black-box approach, outperforming it thanks to the GENERIC formalism, as it adds physical constrains to the prediction that act as an inductive bias. The results show good agreement between our predictions and the synthetic ground truth obtained by CFD for the two examples analysed. However, there are some limitations in the current work that could be improved in the future:
\begin{itemize}
	\item \textbf{Database:} The present work makes use of a synthetic database generated by a CFD tool. However, real data coming from sensors could be used to train a system to work with real-world digital twins. Additionally, the database could be augmented with different geometry cases, improving the generalization to unseen geometries.
	\item \textbf{Integration scheme:} In this work, an Euler integration scheme is used. This is integration scheme is simple, and higher order integration schemes like the midpoint rule, Heun's method or a Runge-Kutta method \cite{RK_NN, GFINNS} could improve the accuracy of the SPNN. This would also allow the network to work with bigger time increments. However, increasing the complexity of the integration scheme would require more forward passes of the neural network for each time step, slowing the training process.
	\item \textbf{Net architecture:} Graph Neural Networks (GNNs) \cite{BATTAGLIA_2018, BRONSTEIN_2017} could be used to take advantage of their unstructured data, in comparison to convolutional neural networks, that require grid-structured information. Thus, GNNs could be applied to real-world applications, e.g., a digital-twin of a system whose sensors are not evenly distributed.
\end{itemize}

\section*{Acknowledgements}

This work was supported by the Spanish Ministry of Science and Innovation, AEI/10.13039/501100011033, through Grant number PID2020-113463RB-C31 and by the Ministry for Digital Transformation and the Civil Service, through the ENIA 2022 Chairs for the creation of university-industry chairs in AI, through Grant TSI-100930-2023-1.

This material is also based upon work supported in part by the Army Research Laboratory and the Army Research Office under contract/grant number W911NF2210271.

This research is also part of the DesCartes programme and is supported by the National
Research Foundation, Prime Minister Office, Singapore under its Campus for Research
Excellence and Technological Enterprise (CREATE) programme.

The authors also acknowledge the support of ESI Group through the chairs at the
University of Zaragoza and at ENSAM Institute of Technology.

\bibliographystyle{unsrt}
\bibliography{references} 

\end{document}